\UseRawInputEncoding
\documentclass[%
 reprint,
 amsmath,amssymb,
 aps,
]{revtex4-2}

\usepackage{graphicx}
\usepackage{dcolumn}
\usepackage{bm}
\usepackage{xcolor}
\usepackage{appendix}

\begin{document}

\preprint{APS/123-QED}

\title{In-Lab High Resolution Mid-infrared Up-conversion Stellar Interferometer Based on Synthetic Long Base-Line}
\author{Zhao-Qi-Zhi Han$^{1,2}$}
\author{Zheng Ge$^{1,2}$}
\author{Wen-Tao Luo$^{3,4}$}
\email{wtluo@ustc.edu.cn}
\author{Yi-Fu Cai$^{4,3}$}
\author{Xiao-Hua Wang$^{1,2}$}
\author{Li Chen$^{1,2}$}
\author{Wu-Zhen Li$^{1,2}$}
\author{Zhi-Yuan Zhou$^{1,2,}$}
\email{zyzhouphy@ustc.edu.cn}
\author{Bao-Sen Shi$^{1,2,}$}
\email{drshi@ustc.edu.cn}

\affiliation{%
$^{1}$CAS Key Laboratory of Quantum Information, University of Science and Technology of China, Hefei, China\\
$^{2}$CAS Center for Excellence in Quantum Information and Quantum Physics, University of Science and Technology of China, Hefei, China\\
$^{3}$Institute of Deep Space Science, Deep Space Exploration Laboratory, Hefei, China\\
$^{4}$Department of Astronomy, School of Physical Sciences, University of Science and Technology of China, Hefei, China}

\date{\today}

\begin{abstract}
Detecting mid-infrared (MIR) radiation has significant astronomical applications, although limited by unsatisfactory MIR detectors. Here we reported on the realization of a MIR up-conversion interferometer based on synthetic long base-line (SLBL) in the laboratory. The experimental system consisted of an interferometer and subsequent up-conversion detection part of mid-infrared signal, which streamlined the structure and enhanced the reliability of the system. By using a tungsten filament lamp as an imitated star, we not only achieved the single target angle resolution of  $1.10 \times 10^{-4} \ \ rad$, but also obtained the field angle resolution of $3.0 \times 10^{-4} \ \ rad$ of double star targets. The angular resolution is in inverse proportion to the length of baseline. The maximum length of simulated baseline in the laboratory is about $3\ cm$. In a Keck Interferometer (KI) liked program, the base line can reach up to $85\ m$ leading to a corresponding angular resolution of $ 3\times 10^{-9} \ \ rad$ (about 1.8mas). The study will offer potential benefits in extending the usage of mid-infrared light in astronomical exploration.

\end{abstract}

\maketitle


\section{\label{sec:level1}Introduction}

The optical/infrared interferometry revolutionized many aspects in observational astronomy, i.e. active galactic neuclei, exoplanet, Galactic Center e.t.c. For instance, the GRAVITY instruments installed at the Very Large Telescope Interferometer (VLTI) \citep{GRAVITY2017AA} with four 8-meter telescopes, can reach up to 200 meter squared effective collecting area and hence microarcsecond level astrometric accuracy. Keck Interferometer (KI) with two 10 meter telescope separating 85 meters away from each other, can achieve 5mas resolution in astrophysical observations \citep{colavita2013PASP}. Such high resolutinon enables the re-evaluation of the super massive black hole mass in the center of active galactic neuclei, resolution of the planetary disc as well as star diameter measurement. 

The mid-infrared (MIR) radiation, ranging from $2.5$ to $25 \ \mu m$, holds notable value in the fields of spectroscopy and imaging for astronomical studies. It is particularly useful for observing the formation or evolution of active galactic nuclei, young stellar objects, and planetary systems\ \cite{cody2014aj,marigo2017aj,wang2019aj,quanz2022aa,thilker2023ajl}. To measure abundant information of celestial bodies using MIR light, high sensitive MIR detectors with extremely high angular resolution are required. The atmospheric transparent windows of $3-5 \ \mu m$ and $8-14\ \mu m$ enables the ground-based interferometry observations with less scattering effect due to the long wave length of MIR photons. 

Traditional MIR semiconductor detectors have some disadvantages in terms of detection sensitivity, noise level, response time, and cost compared to detectors working in visible or near-infrared band. Common used detectors working in MIR band, based on narrow bandgap materials, such as mercury cadmium telluride (MCT) and indium arsenide (InAs), as well as superconducting detection technologies, have achieved high sensitivity and quantum efficiency. However, they heavily rely on the deep cooling to reduce dark noise which limit their applications in some practical environments. In order to achieve high angular resolution of observing a stellar object, the stellar interferometers based on synthesizing long baseline are used. By increasing the distance between two receivers to increase the baseline length, the resolution can be improved by orders of magnitude.

The combination of the synthetic long baseline stellar interferometer and the up-conversion detection (UCD) is an effective solution to detect the MIR radiation with high sensitivity and high angle resolution. Using UCD can convert MIR signal light into visible/near-infrared band through sum-frequency generation (SFG) process, then the high-performance detectors based on broadband gap materials, such as silicon, can be used for detection\ \cite{han2023adi}. UCD offers fast response speed and room temperature operation, which has been applied in fields like quantum information or optical imaging\ \cite{mancinelli2017nc,ge2022cpb,ge2023apn,ge2023prapplied,dam2012np}. The method also achieved some progress in astronomy both in laboratory and practical applications in the H-band\ \cite{ceus2012mnrasl,gomes2014prl,darre2016pr,lehmann2018mnras} and is gradually developing towards the MIR band\ \cite{szemendera2016mnras,lehmann2019mnras,magri2020mnras}.

This article presents a detailed study on the long baseline Michelson stellar interferometer with UCD under laboratory conditions. The relationship between the linearity and field angle of the interference curve of the experimental configuration has been theoretically analyzed. A calculation method for the conversion efficiency of the SFG process is also provided. The size of a series of simulated stars is measured in the laboratory, along with the field angle of star, radius, and distance between two stars. The maximum spacing between the two observers of the interferometer in our laboratory is only $34\ mm$, an angular resolution of $1.10\times10^{-4}\ rad$ is obtained at a distance between the star and observer of $4.98\ m$, with an average error rate of about $3 \ \% $ and a signal-to-noise ratio (SNR) of over 3. The center wavelength used for detection is $3.27\ \mu m$.

Compared to previous works, an interference-first structure followed by up-conversion detection is used. This approach offers several benefits, including the reduction of system complexity and the elimination of the need for multiple up-conversion systems. Additionally, the use of a single nonlinear crystal allows for convenient adjustment of bandwidth (corresponding coherence length), central wavelength, and quantum efficiency to meet the requirements of various application scenarios. In addition, the system's ability to resist jitter is enhanced by using MIR light with longer wavelengths for interference, which is beneficial for more accurate target measurement.

\section{Theoretical analysis}
\subsection{Interference curve and visibility}
For an ideal Michelson Stellar interferometer with infinite long coherence length, its intensity can be expressed as \cite{scully1999book}:
\begin{align}
    \bm{I}=2\bm{I}_0\{1+\cos[\frac{(\vec{\bm{k}}+\vec{\bm{k}}')\cdot(\vec{\bm{r}}_1-\vec{\bm{r}}_2)}{2}] \nonumber \\
    \cos(\frac{\pi\alpha\bm{d}}{\lambda})\}
\end{align}
where $\bm{I}_0$ is the intensity of a single path of the interferometer; $\vec{\bm{k}}$ and $\vec{\bm{k}}'$ $(\lvert\vec{\bm{k}}\rvert=\lvert\vec{\bm{k}}'\rvert=\frac{2\pi}{\lambda})$ are wave vectors of radiation from both ends of the celestial body; $\vec{\bm{r}}_1$ and $\vec{\bm{r}}_2$ are the displacements from the center of the celestial body to the two slits of the interferometer; $\bm{d}=\lvert\vec{\bm{r}}_1-\vec{\bm{r}}_2\rvert$ is the distance between two slits; $\alpha$ is the full field angle of the celestial body to the interferometer; $\lambda$ is the working wavelength of the interferometer. However, due to the spatial size of the double slits of the interferometer and the celestial body, it can disrupt the spatial coherence of the detected radiation. The accurate form of its intensity of the interference curve can be written as \cite{born1987book}:
\begin{equation}
 \bm{I}=2\bm{I}_0[1+\gamma\cos(\Delta\varphi)]
\end{equation}
where $\gamma=\bm{k}\mathrm{sinc}(\frac{\pi\alpha\bm{d}}{\lambda}))$ is the interference visibility, $\bm{k}$ is the interference coefficient which is affected by system stability, slit width, wavelength bandwidth, etc. When the distance between two slits is increased so that the phase difference between the two channels of the interferometer exactly reaches $\pi$, the interference visibility $\gamma$ becomes 0, and there is no interference phenomenon in the interferometer, the corresponding full field angle of a celestial body can be calculated as:
\begin{equation}
    \alpha=\frac{\lambda}{\bm{d}_0}
\end{equation}
Where, $\bm{d}_0$ is the distance between two slits when interference disappears.
\subsection{Up-conversion detection}
The quantum theory of the sum frequency generation (SFG) pumped by continuous wave in a second-order nonlinear crystal is shown as follows. Under the condition of undepleted pump, an SFG photon at frequency $(\omega_{SFG}=\omega_p+\omega_s)$ is produced by the simultaneously annihilation of a signal photon at frequency $\omega_s$ and a pump photon at $\omega_p$. Under the phase matching condition, the Hamiltonian of the SFG can be written as\cite{kumar1990ol}:
\begin{equation}
    \widehat{\bm{H}}_{TW\!M}=\bm{i}\hbar\xi\widehat{\bm{a}}_s\widehat{\bm{a}}^\dag_{SFG}+\bm{H.c.}
\end{equation}
where $\widehat{\bm{a}}_i$ and $\widehat{\bm{a}}^\dag_i$ $(i=s,SFG)$ represent the annihilation and creation operators of the signal and SFG photons, respectively; $\xi=\bm{g}\bm{E}_p$ is a constant, which is proportional to the amplitude of the strong pump $\bm{E}_p$ and the coefficient of second-order polarization tensor $\bm{g}$. In the Heisenberg’s picture, 
\begin{equation}
  \begin{array}{c}
{\widehat a_s}\left( L \right) = {\widehat a_s}\left( 0 \right)\cos \left( {g{E_p}L} \right) - {\widehat a_{SFG}}\left( 0 \right)\sin \left( {g{E_p}L} \right)\\
{\widehat a_{SFG}}\left( L \right) = {\widehat a_{SFG}}\left( 0 \right)\cos \left( {g{E_p}L} \right) + {\widehat a_s}\left( 0 \right)\sin \left( {g{E_p}L} \right)
\end{array}
\end{equation}
Here $L$ is the length of the nonlinear crystal. When $g{E_p}L = \frac{\pi }{2}$, ${\widehat a_{SFG}}\left( L \right) = {\widehat a_s}\left( 0 \right)$, representing that all the signal photons are converted into SFG photons. The quantum efficiency of SFG can be obtained by comparing the numbers of SFG photons to input signal photons, $\eta  = {\raise0.7ex\hbox{${{{\widehat N}_{SFG}}\left( L \right)}$} \!\mathord{\left/
 {\vphantom {{{{\widehat N}_{SFG}}\left( L \right)} {{{\widehat N}_s}\left( 0 \right)}}}\right.\kern-\nulldelimiterspace}
\!\lower0.7ex\hbox{${{{\widehat N}_s}\left( 0 \right)}$}} = {\sin ^2}\left( {g{E_p}L} \right)$ where ${\widehat N_j} = \left\langle {\widehat a_j^\dag {{\widehat a}_j}} \right\rangle $ and ${\widehat N_{SFG}}\left( 0 \right) = 0$. The general expression for the conversion efficiency can be defined as\cite{albota2004ol}:
\begin{equation}
    \eta  = {\sin ^2}\left( {\frac{\pi }{2}\sqrt {\frac{P}{{{P_{\max }}}}} } \right)
\end{equation}
Where ${P_{\max }}$ is the pump power required when the quantum conversion efficiency reaches $100\%$.
\section{Experiment Setup}
\begin{figure*}
  \centering\includegraphics[width=110mm]{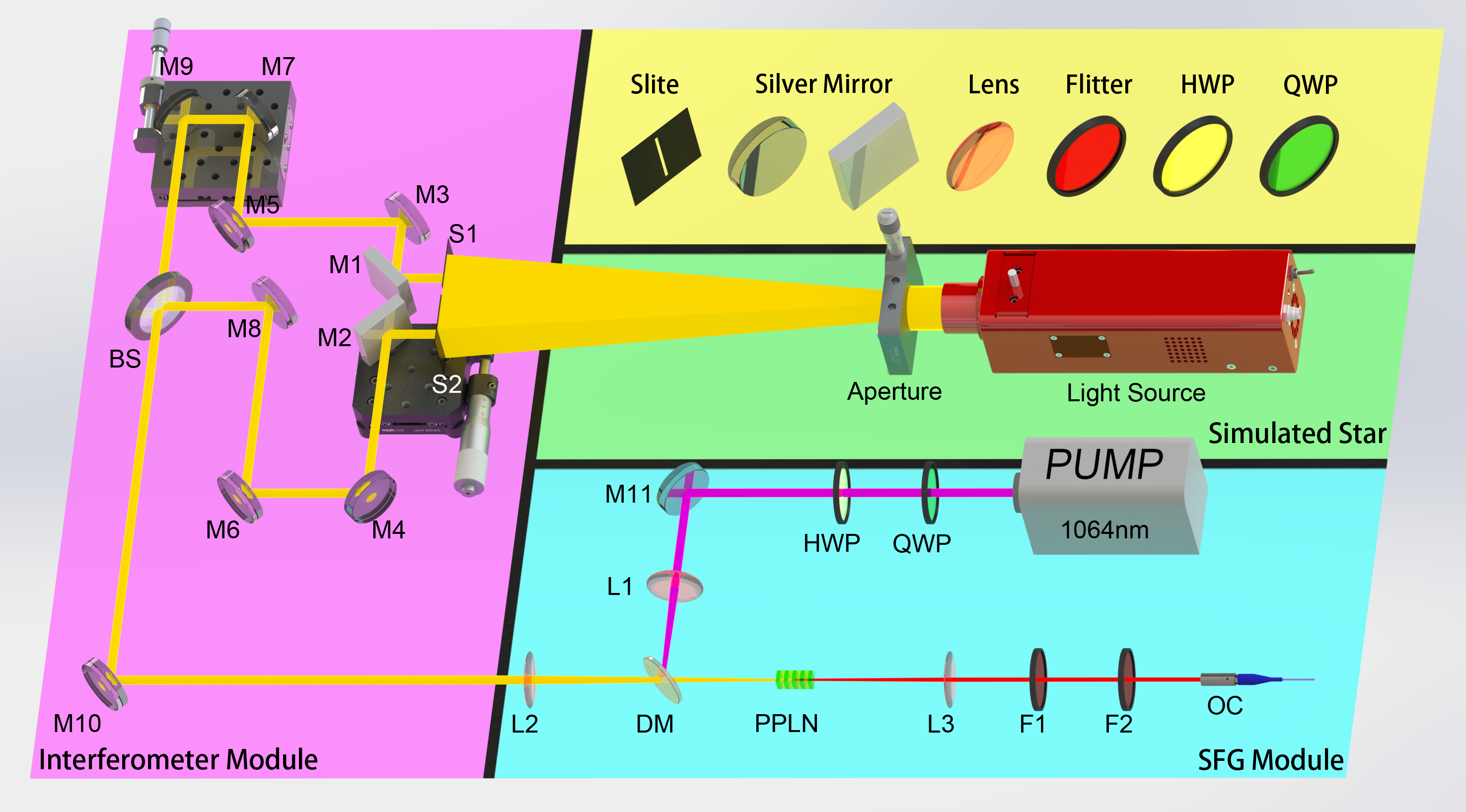}
  \caption{Schematic diagram of the experimental setup. S terms: the slits; L terms: the lenses; M terms: the sliver mirrors; F terms: the flitters; HWP: half-wave plate; QWP: quarter-wave plate; DM: dichromatic mirror; PPLN: periodically poled lithium niobate crystal; OC: optical coupler.}
  \label{fig:boat1}
\end{figure*}
The experimental setup is presented in \textbf{Fig. \ref{fig:boat1}}, which consists of three parts: MIR stimulated star source, Michelson stellar interferometry and SFG module. The green area shows the MIR stimulated star source. Where, radiation from a tungsten halogen light source (SLS202L/M, Thorlabs), is collimated and input through an aperture acting as radiation from a celestial body. A band-pass filter with a bandwidth of $2000\ nm$ centered at $4000\ nm$ has been inserted in the tungsten halogen light to filter out visible light. The celestial body can be regarded as a point light source due to the fact of over long distance from the earth. A Michelson stellar interferometry is placed in the pink area in \textbf{Fig. \ref{fig:boat1}}. We use two $1\ mm$ slits to act as the receiving end of the Michelson stellar interferometer. S2 and M2 are positioned on the displacement plate, allowing for adjustment of the distance d between the two slits. M7 and M9 are placed on another displacement plate as a delay to adjust the path length difference between two arms of the interferometer. The path difference in the experiment should be less than the coherence length of the radiated light. Two light beams intersect and interfere on the beamsplitter (BS) (BP145B4, Thorlabs), and then pass into the UCD system via mirror M10. The SFG module is placed in the blue area in  \textbf{Fig. \ref{fig:boat1}} and is used for UCD. The size of the type-0 phase matched periodically poled lithium niobate (PPLN) crystal for SFG is $0.5\ mm\ \times \ 0.5\ mm\ \times \ 40\ mm$, poling period is of $22.4\ \mu m$. The signal MIR photons, after passing through L1 with a focal length of $75\ mm$, are injected into the crystal. The CW pump light outputs from a Yb-doped fiber amplifier at $1065\ nm$. The lens L2 has a focal length of $150\ mm$. Ultimately, the upconverted photons at $803\ nm$ are collimated, filtered and coupled into a single-mode fiber, which is fed to the silicon single photon avalanche photodiode (SPAD) (SPCM-AQRH-14-FC, Excelitas) for detection. The lens L3 has a focal length of $150\ mm$ and the filter system consists of a dichromatic mirror and a band-pass filter centered at $800\ nm$ and a bandwidth of $10\ nm$.

\section{Result}
\subsection{Interference intensity and visibility}

\begin{figure*}
  \centering\includegraphics[width=\textwidth]{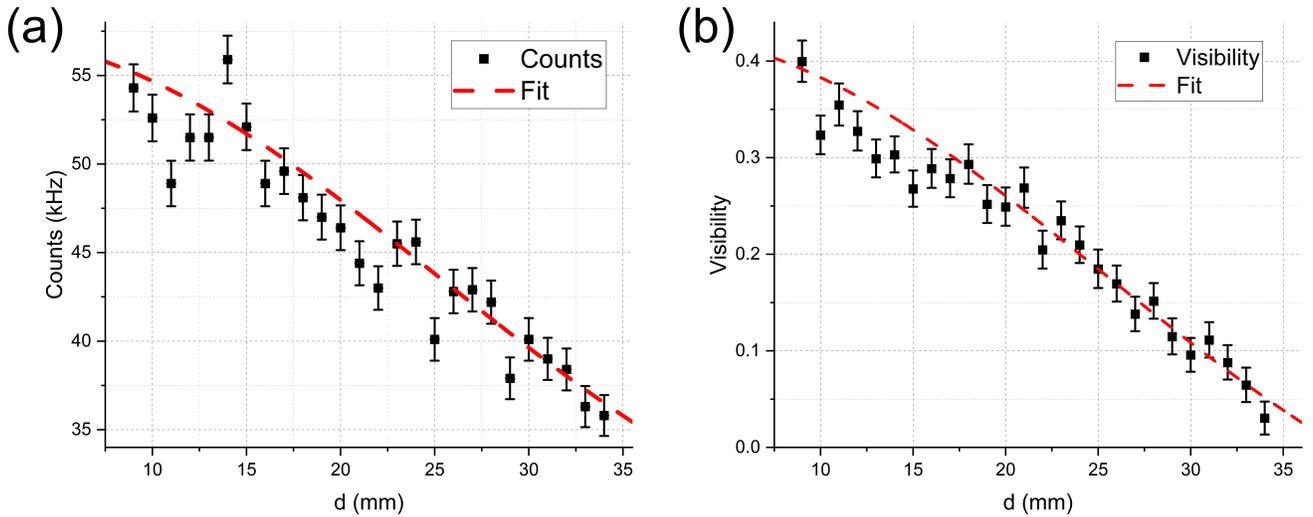}
  \caption{Interference characteristics of the system when a $0.45\ mm$ slit is used as the aperture of MIR source. (a) Photon counts rate of maximum interference intensity at different distance d between double slits. The noise photon counts have been subtracted. (b) Interference visibility at distance d between double slits. The experimental results are represented by the black square, while the fitting curve are represented by the red dashed line in both two images.} 
  \label{fig:boat2}
\end{figure*}

We first evaluated the performance of the synthetic long base-line stellar interferometer combined with the MIR up-conversion. A $0.45\ mm$ wide slit is placed in front of the tungsten halogen light source as an aperture to measure the maximum intensity and visibility of the interference against the distance between S1 and S2. When S2 moves, the photon counts of either path remain stable. The positions of the silver mirrors M7 and M9 also need to be adjusted according to d to ensure that the difference between two arms is less than the coherent length. The experimental results are shown in \textbf{Fig. \ref{fig:boat2}}. As the distance between the two slits increases, the maximum interference intensity and visibility of the interferometer decreases. At the maximum adjustable distance between double slits of $34\ mm$, weak interference is still observable with the visibility of only $(3.02\ \pm\ 0.82)\ \%$. The signal photon counts is $35.3\ kHz$, slightly higher than the sum of the counts of either path of $35.1\ kHz$. During the measurement process, the average noise photon count is $10.0\ kHz$, and the signal-to-noise ratio (SNR) at the point of interference maximum is greater than 3.5. The two experimental results are fitted by using the equation 
$\gamma  = k\sin c\left( {\frac{{\pi \alpha d}}{\lambda }} \right) $ function, where $0.0832\ mm^{-1}$ and $0.0823\ mm^{-1}$ are the fitting results of $\pi \alpha /\lambda$, respectively. The calculated theoretical coefficient is $0.0868\ mm^{-1}$, which is slightly greater than our experimental results.

\subsection{Measurement of simulated stellar size}
\begin{figure*}
  \centering\includegraphics[width=\textwidth]{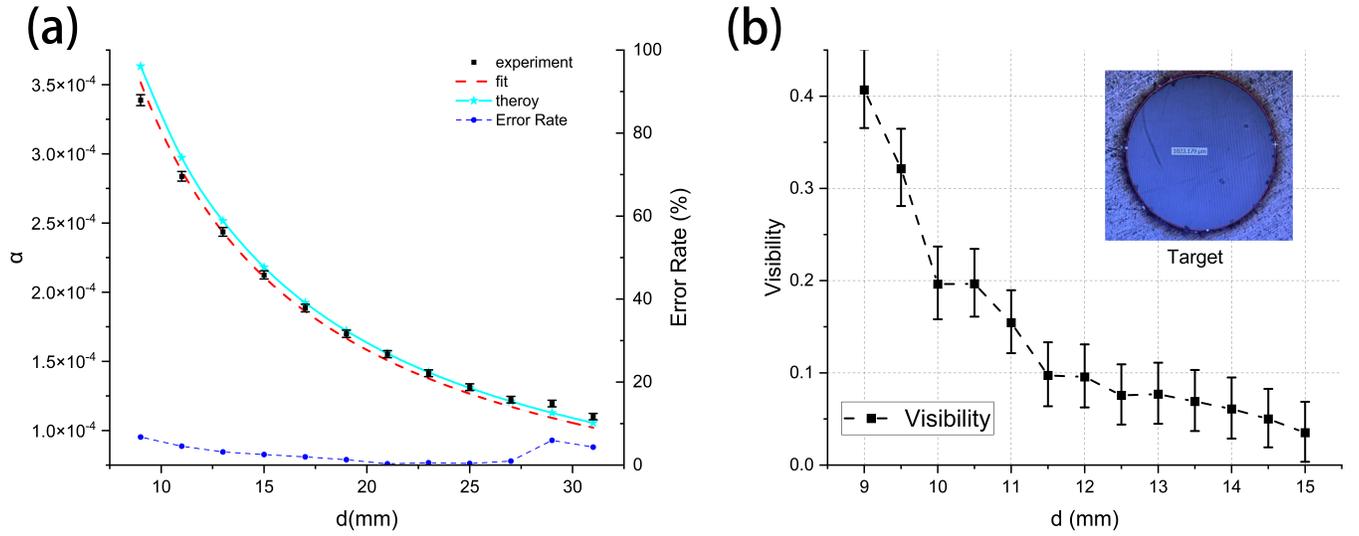}
  \caption{Measurement results of simulated stars prepared with different apertures. (a) The field angle measured by different double slits distances by using an adjustable width slit as the aperture, including experimental results (black points), theoretical curves (sky blue line), fitting curves (red dashed line), and error rates (blue-black points). (b) Measurement of interference visibility and field angle for a circular small hole as target. The microscopic image and size of the small hole are also provided in the upper right corner.} 
  \label{fig:boat3}
\end{figure*}
Then we use various apertures as simulated stars and measure their corresponding field angles by observing the positions where interference disappears, rather than by measuring the maximum interference intensity and visibility at different double slit distances. This method is both convenient and accurate. In this case, a two-dimensional adjustable slit is used in front of the tungsten halogen source as the stimulated star. The slit's length and width can be changed between $0$ and $10\ mm$. We measure the width of the slit when interference disappears while distance between the double slits is kept unchanged, the length of the slit is consistently set to $10\ mm$ to enable the receiving end to collect more signal photons. The adjustable slit is positioned $4.98\ m$ away from the double slits and the center wavelength of the conversion bandwidth of the PPLN crystal is $3.27\ \mu m$. For the double slits distance, we selected 12 points as ${d_0}$ between $9\ mm$ and $31\ mm$ by each step of $2\ mm$ to measure the corresponding stimulated stellar field angle, which change from $(3.38\ \pm\ 0.04)\ \times\ 10^{-4}\ rad$ to $(1.10\ \pm\ 0.02)\ \times\ 10^{-4}\ rad $. The calculation of the corresponding field angles for different values of double slits distance and the angles corresponding to the selected distance in experiment are shown in \textbf{Fig. \ref{fig:boat3}(a)} by the sky blue lines and points. The measurement results show an average error rate of only $2.73\ \%$. Additionally, a function $\alpha  = {\raise0.7ex\hbox{${{\lambda _f}}$} \!\mathord{\left/
 {\vphantom {{{\lambda _f}} d}}\right.\kern-\nulldelimiterspace}
\!\lower0.7ex\hbox{$d$}}$ is used to numerically fit the measurement results, which is also shown in \textbf{Fig. \ref{fig:boat3}(a)} as the red dash line. The fitted coefficient ${\lambda _f}$ is $3.17\ \mu m$, slightly smaller than the center wavelength selected in the experiment, with an error rate of only $3.06\ \%$. The circular hole target is a simulated target that closely resembles the actual celestial body. A small hole with a diameter of $1023.179\ \mu m$ is measured. The circular hole permits the more few light passing through compared to the adjustable slit, resulting in a significant reduction in the counts of signal photons that can be received and detected, which leads a subsequent decrease in the SNR. During this measurement process, the average noise count is $10.7\ kHz$, and the minimum total photon count is $16.6\ kHz$, resulting in the SNR of only 0.55. The interference visibility is measured by moving the slit S2 and sliver mirror M2 by a step of $1\ mm$ at each time. When the distance between double slits reaches $15\ mm$, the lowest interference visibility is found to be $0.35\ \%$. It is believed that this distance ${d_0}$ is the location where interference disappears. In this experiment, the field angle corresponding to the circular hole is measured to be $2.18\ \times\ 10^{-4}\ rad$, which is slightly larger than the actual angle of $2.05\ \times\ 10^{-4}\ rad$. 

This system can also be applied to the field angle measurement of two stars to interferometer. We drill two pinholes with a spacing of $1.53\ mm$ on the aperture, and the diameter of the pinholes is much smaller than the spacing. We observe that the interference phenomenon disappears when the double slits distance reaches $10.9\ mm$, at this time, the field angle measured is about $3\ \times\ 10^{-4}\ rad$.

\subsection{Future astronomical application}
\begin{figure}
  \centering\includegraphics[width=9cm,height=7cm]{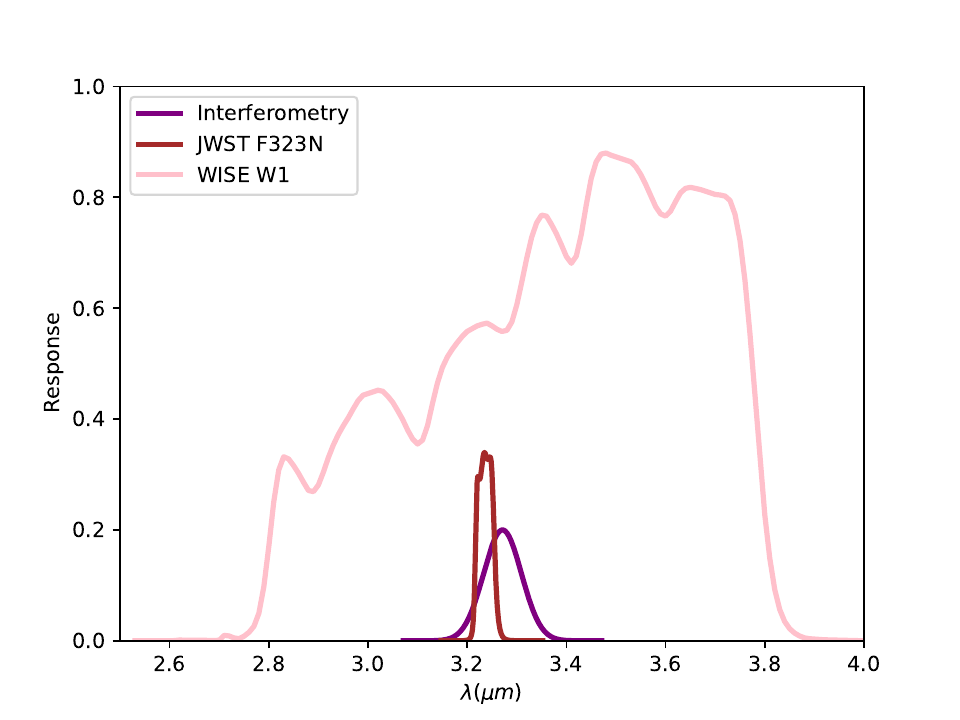}
  \caption{This figure shows the comparison among the MIR filters of three instruments. The WISE W1 filter(pink solid line), JWST F323N filter(brown solid line) and the response of our instrument(purple solid line). } 
  \label{fig:filters}
\end{figure}

The Event Horizon Telescope (EHT) has obtained the 25 microarcsecond resolution of the central super massive black hole of M87 at 1.3mm wavelength \cite{EHT2019ApJ} which is the highest resolution so far that can image the very close vicinity of the super massive black hole. This well studied source has also been observed by infrared instruments with 0.2-0.3 arcsecond resolution based on HST \cite{Prieto2016MNRAS}. The infrared photometric and structural observations from the disc can provide valuable information for the AGN mechanism. The JWST F323N narrow band can achieve a resolution of 0.1 arcsecond \cite{Rieke2023PASP}, the WISE telescope with W1 filter has the resolution of 2 arcsecond \cite{Wright2010AJ} with broader response curve. Assuming a Keck-like instrument separating 85 meters installing our interferometric setups, and the filter response is shown in Fig.~\ref{fig:filters} as purple line which is broader than our original set. This can be achieved by applying shorter length crystal \cite{Ge2024SciAdv}. The WISE W1 filter (pink line) and JWST F323N (brown line) are also plotted as comparison.

\section{Conclusion}

In summary, we set up a first-order amplitude SLBL interferometer in the laboratory, by which we can measure the field angle of the celestial bodies with the aid of nonlinear frequency up-conversion technique. Our system provides an effective alternative to detect and measure the mid-infrared signals emitted by celestial bodies with the advantages of room-temperature operation and real-time conversion. We measure the target with a corresponding field angle of $1.10\times10^{-4} \ rad$ at a distance of $4.98\ m$ by moving the double slits spacing and measuring interference visibility with the help of the SLBL. On the contrary, the minimum angle that is measured using a $1\ mm$ slit at the same distance is only $4\times10^{-3}\ rad$, is more than one order lower than the SLBL stellar interferometer with up-conversion. We have verified that there is an inverse proportional relationship between the resolution of the interferometer and the baseline length. In the actual measurement process, the resolution can be further improved 2 or even more order of magnitude by increasing the spacing between telescopes to lengthen the SLBL. The complexity of the system is simplified by interference first and then up-conversion, compared to Ref.\ \cite{magri2020mnras}. Up-conversion acts also as a good narrowband filter, which will further improve the equivalent coherence length of mid-infrared signal. The coherence length of our experimental configuration is mainly determined by the conversion bandwidth of the PPLN crystals. Although natural light with an extremely wide bandwidth interferes on BS, the limited bandwidth that PPLN can convert will greatly increase the coherence length of the system. It can be theoretically calculated that central wavelength of the PPLN crystal is $3271.16\ nm$, and the conversion bandwidth is $4.28\ nm$ at $25\ ^\circ C$, resulting in a corresponding coherence length of $2.56\ mm$. Altering the length of the crystal can change its acceptance bandwidth to adjust the coherence length. The center wavelength of the SFG process can be adjusted by changing temperature of the crystal and the wavelength of the pump. Additionally, the wavelength of mid-infrared light is longer. The same size of jitter causes less phase change in the MIR interferometer, which leads better anti-jitter characteristics. As for interference visibility, the maximum interference visibility of the experimental configuration is affected by two main factors. Firstly, the splitting ratio of BS is not well controlled at $1:1$ due to the immaturity of mid-infrared devices. In the experiment, we used the BS with a splitting ratio of nearly $3:1$ at $3.27\ \mu m$ for p-polarized light, which significantly affects interference visibility. The width of the double slits at the receiving end is the second factor affecting interference visibility. Increasing the slit width will decrease interference visibility. The Michelson interferometer in the mid-infrared band with up-conversion detection will further enhance its application in star target recognition and provide a more sensitive and accurate detection method. This work provides a new platform for MIR band in astronomical interferometer. Future work will continue to investigate its practical application, the up-conversion interferometers of other new systems and their regulatory characteristic.

\par
\begin{acknowledgments}
We would like to acknowledge the support from the National Key Research and Development Program of China (2022YFB3607700, 2022YFB3903102), National Natural Science Foundation of China (NSFC) (11934013, 92065101, 62005068), Innovation Program for Quantum Science and Technology (2021ZD0301100), the Space Debris Research Project of China (No. KJSP2020020202), and USTC Research Funds of the Double First-Class Initiative.
\end{acknowledgments}

\appendix
\section{Quantum Efficiency of SFG}
We first evaluate the quantum efficiency of this SFG under different pump power. We measure the visible beam power output by a MIR laser and a strong pump laser through the nonlinear crystal after up-conversion. The conversion efficiency of SFG can be written as
\begin{subequations}
\begin{eqnarray}
&{\eta _{power}}&=\frac{{{P_{visible}}}}{{{P_{mir}}}} \times 100\%, \label{appa}
\\
&{\eta _{quantum}}&=\frac{{{\nu _{mir}}}}{{{\nu _{visible}}}}{\eta _{power}} = \frac{{{\lambda _{visible}}}}{{{\lambda _{MIR}}}}{\eta _{power}}, \label{appb}
\end{eqnarray}
\end{subequations}
Where, ${\eta _{power}}$ and ${\eta _{quantum}}$ represent power efficiency and quantum efficiency, respectively. ${P_{mir}}$ is the input power of MIR light, and ${P_{visible}}$ is the output power of visible light. ${\nu _{i}}$ and ${\lambda _i}\left( {i = mir,visible} \right)$ represent the frequency and wavelength of two light, respectively.

The MIR classical source is generated by a difference-frequency generation (DFG) process with power of $0.18\ mW$ at $3270\ nm$, which is much smaller than power of pump at $1065\ nm$. This can meet the approximation of nondepleting pump, which means that the pump power in SFG does not change. The maximum power of the CW pump output from the Yb-doped fiber amplifier is $15\ W$.  We measure that the output power of visible light at $803\ nm$ is $0.089\ mW$ under the condition of pump power of $15\ W$, achieving a power conversion efficiency of $49.4\%$ or a quantum conversion efficiency of $12.1\%$.

Within the range of pump light power less than $15\ W$, there is a linear relationship between visible light power and pump power, which means that our pump power is much smaller than ${P_{max}}$ in Eq.(6) in the main text. Therefore, the quantum conversion efficiency per Watt of pump power can be calculated to be $0.8\%\cdot {W^{ - 1}}$. The test result in this condition can be applied to the up-conversion detection of MIR light in the single photon level in our experiment for prediction and calculation. 

Our subsequent experiments are conducted at a pump power of $5\ W$, at which the laser output power can remain stable for a long time, corresponding to a quantum conversion efficiency of $4\%$ inside the crystal.

\section{Bandwidth of Light Source and SFG}
Tungsten halogen light source is a form of incandescent lamp, which is filled with halogen gas inside that can suppress the sublimation of tungsten wire to heat up to around a couple of thousand of Kelvin. It can be approximated as a black body radiation source and is very suitable for simulating stellar luminescence.

The Tungsten halogen light source we use (SLS202L/M, Thorlabs) generates output from $450\ nm$ to $5500\ nm$. A band-pass filter with a bandwidth of $2000\ nm$ centered at $4000\ nm$ ($\#11-998$, Edmund) has been inserted in the tungsten halogen light to filter out visible light and ensure that the emitted light retains the thermal radiation field characteristics. Therefore, the actual output of the MIR source has a wavelength range from $3000\ nm$ to $5000\ nm$. The output power decreases with wavelength, which approximately satisfies the black body radiation law.

As for the bandwidth of the up-conversion process, the output power of SFG process is related to phase mismatch and crystal length, which can be expressed as
\begin{equation}
P \propto \mathrm{sinc^2}\left( {{\raise0.7ex\hbox{${\Delta kL}$} \!\mathord{\left/
 {\vphantom {{\Delta kL} 2}}\right.\kern-\nulldelimiterspace}
\!\lower0.7ex\hbox{$2$}}} \right)
\end{equation}
where $\Delta k = 2\pi \left( {\frac{{{n_{vis}}}}{{{\lambda _{vis}}}} - \frac{{{n_{nir}}}}{{{\lambda _{nir}}}} - \frac{{{n_{mir}}}}{{{\lambda _{mir}}}} - \frac{1}{\Lambda }} \right)$ is the phase mismatch; ${n _{i}}$ and ${\lambda _i}\left( {i = vis,nir,mir} \right)$ represent the refractive index and wavelength of visible light converted by SFG, near-infrared(NIR) pump and MIR signal respectively; ${\Lambda}$ represents the length of crystal polarization period; $L$ is the length of crystal. The full width at half maximum(FWHM) of the SFG in our experiments can be theoretically calculated as $4.28\ nm$, resulting in a corresponding coherence length of $2.50\ mm$. The coherence length can be measured by moving the displacement tables where M7 and M9 are located in our experiment, which can reflect our conversion bandwidth. In the experiment, when we move the displacement table $1.2\ mm$, the interference disappeared, corresponding to a coherence length of $2.4\ mm$, which is consistent with our theoretical calculation result.

In summary, the actual detected MIR bandwidth is determined by our up-conversion process, which is approximately $4\ nm$, corresponding to an $R = {\raise0.7ex\hbox{$\lambda $} \!\mathord{\left/
 {\vphantom {\lambda  {\Delta \lambda }}}\right.\kern-\nulldelimiterspace}
\!\lower0.7ex\hbox{${\Delta \lambda }$}} \sim 800$. At present, the spectral width we are using is relatively small. In fact, we can simply increase the conversion bandwidth by reducing the crystal length while ensuring a reasonable coherence length, thereby further improving the energy utilization efficiency.

\section{Brightness $\&$ Photon Counts}
For our entire up-conversion interferometer system, the actual observed net photon count rates is only $5.9\ kHz$ when measuring the diameter of the stimulated stellar body which is generated by using a single circular hole. Our entire system has a detection efficiency of $1.1\%$ (with $4\%$ up-conversion quantum efficiency when the pump power is 5W, $58\%$ visible photons coupling efficiency, $62\%$ detect efficiency for our detector at $803\ nm$, $18\%$ loss introduced by reflection and transmission mirrors, and about $6\%$ of photons passing through another path at maximum position caused by the 3:1 splitting ratio of BS), resulting in about $5.35 \times 10^{5}$ MIR photons per second in the conversion bandwidth coming from the stimulated stellar body can be received by our system. The total energy corresponding to these photons is about $32\ fW$, and the corresponding energy density is $7.6\ fW\cdot {nm^{ - 1}}$. 
Our average noise count rate in the experiment is $10.7\ kHz$, which mainly comes from the photons introduced by our laboratory environment and pump light. We only used one $750\ nm$ long-pass filter and two $1000\ nm$ short-pass filters at the detection end. Therefore, these noise photons can be reduced by selecting more suitable narrow band-pass filter and packaging the system. 

For the brightness of our MIR light source, its spectral energy density output near $3270\ nm$ is approximately $0.22\ \mu W\cdot {nm^{ - 1}}$. The transmittance of the filter (4000-2000 band-pass filter) we use is about $85\%$ near $3270\ nm$. And we use its matching kit to collimate the output light of the lamp, and its all band output power will be reduced from $700\ mW$ for free space output to $15\ mW$ for collimated output with a diameter of about 10mm, which can be found in its manual. We assume that the losses at each wavelength are consistent, but in reality, due to the lens material, longer wavelengths will suffer more losses. In order to allow more power to pass through our $1\ mm$ diameter circular hole, we focuse the beam using a CaF2 lens with a focal length of $50\ mm$ and placed the hole on the back focal plane of the lens. The diameter of the beam on the back focal plane of the lens is $4\ mm$, which means only about $\frac{1}{{16}}$ of the energy from source after collimation kit can pass through the hole. MIR light will reach our receiving end after being transmitted about $5\ m$ through the back focal plane of the CaF2 lens, with a distance of 100 times our focal length. Through simple geometric relationships, it can be estimated that the energy at our receiving end will be uniformly distributed within a circle with a diameter of approximately $1000\ mm$. The circular hole only changes the size of the light spot distribution, but does not alter the energy density distribution in the receiver plane. Our receiving ends are two $1\ mm$ slits with a length of $25\ mm$, resulting in that we can only utilize about $6.4 \times 10^{-6}$ of the energy from the hole. In summary, we can estimate that the spectral density we can receive is about $16\ fW\cdot {nm^{ - 1}}$, which is consistent with the brightness derived from our experimental results.

\nocite{*}

\bibliography{apssamp}

\providecommand{\noopsort}[1]{}\providecommand{\singleletter}[1]{#1}%
\begin{thebibliography}{29}%
\makeatletter
\providecommand \@ifxundefined [1]{%
 \@ifx{#1\undefined}
}%
\providecommand \@ifnum [1]{%
 \ifnum #1\expandafter \@firstoftwo
 \else \expandafter \@secondoftwo
 \fi
}%
\providecommand \@ifx [1]{%
 \ifx #1\expandafter \@firstoftwo
 \else \expandafter \@secondoftwo
 \fi
}%
\providecommand \natexlab [1]{#1}%
\providecommand \enquote  [1]{``#1''}%
\providecommand \bibnamefont  [1]{#1}%
\providecommand \bibfnamefont [1]{#1}%
\providecommand \citenamefont [1]{#1}%
\providecommand \href@noop [0]{\@secondoftwo}%
\providecommand \href [0]{\begingroup \@sanitize@url \@href}%
\providecommand \@href[1]{\@@startlink{#1}\@@href}%
\providecommand \@@href[1]{\endgroup#1\@@endlink}%
\providecommand \@sanitize@url [0]{\catcode `\\12\catcode `\$12\catcode `\&12\catcode `\#12\catcode `\^12\catcode `\_12\catcode `\%12\relax}%
\providecommand \@@startlink[1]{}%
\providecommand \@@endlink[0]{}%
\providecommand \url  [0]{\begingroup\@sanitize@url \@url }%
\providecommand \@url [1]{\endgroup\@href {#1}{\urlprefix }}%
\providecommand \urlprefix  [0]{URL }%
\providecommand \Eprint [0]{\href }%
\providecommand \doibase [0]{https://doi.org/}%
\providecommand \selectlanguage [0]{\@gobble}%
\providecommand \bibinfo  [0]{\@secondoftwo}%
\providecommand \bibfield  [0]{\@secondoftwo}%
\providecommand \translation [1]{[#1]}%
\providecommand \BibitemOpen [0]{}%
\providecommand \bibitemStop [0]{}%
\providecommand \bibitemNoStop [0]{.\EOS\space}%
\providecommand \EOS [0]{\spacefactor3000\relax}%
\providecommand \BibitemShut  [1]{\csname bibitem#1\endcsname}%
\let\auto@bib@innerbib\@empty
\bibitem [{\citenamefont {{GRAVITY Collaboration}}\ \emph {et~al.}(2017)\citenamefont {{GRAVITY Collaboration}}, \citenamefont {{Abuter, R.}}, \citenamefont {{Accardo, M.}}, \citenamefont {{Amorim, A.}}, \citenamefont {{Anugu, N.}}, \citenamefont {{Ávila, G.}}, \citenamefont {{Azouaoui, N.}}, \citenamefont {{Benisty, M.}}, \citenamefont {{Berger, J. P.}}, \citenamefont {{Blind, N.}}, \citenamefont {{Bonnet, H.}}, \citenamefont {{Bourget, P.}}, \citenamefont {{Brandner, W.}}, \citenamefont {{Brast, R.}}, \citenamefont {{Buron, A.}}, \citenamefont {{Burtscher, L.}}, \citenamefont {{Cassaing, F.}}, \citenamefont {{Chapron, F.}}, \citenamefont {{Choquet, É.}}, \citenamefont {{Clénet, Y.}}, \citenamefont {{Collin, C.}}, \citenamefont {{Coudé du Foresto, V.}}, \citenamefont {{de Wit, W.}}, \citenamefont {{de Zeeuw, P. T.}}, \citenamefont {{Deen, C.}}, \citenamefont {{Delplancke-Ströbele, F.}}, \citenamefont {{Dembet, R.}}, \citenamefont {{Derie, F.}}, \citenamefont {{Dexter, J.}}, \citenamefont {{Duvert, G.}},
  \citenamefont {{Ebert, M.}}, \citenamefont {{Eckart, A.}}, \citenamefont {{Eisenhauer, F.}}, \citenamefont {{Esselborn, M.}}, \citenamefont {{Fédou, P.}}, \citenamefont {{Finger, G.}}, \citenamefont {{Garcia, P.}}, \citenamefont {{Garcia Dabo, C. E.}}, \citenamefont {{Garcia Lopez, R.}}, \citenamefont {{Gendron, E.}}, \citenamefont {{Genzel, R.}}, \citenamefont {{Gillessen, S.}}, \citenamefont {{Gonte, F.}}, \citenamefont {{Gordo, P.}}, \citenamefont {{Grould, M.}}, \citenamefont {{Grözinger, U.}}, \citenamefont {{Guieu, S.}}, \citenamefont {{Haguenauer, P.}}, \citenamefont {{Hans, O.}}, \citenamefont {{Haubois, X.}}, \citenamefont {{Haug, M.}}, \citenamefont {{Haussmann, F.}}, \citenamefont {{Henning, Th.}}, \citenamefont {{Hippler, S.}}, \citenamefont {{Horrobin, M.}}, \citenamefont {{Huber, A.}}, \citenamefont {{Hubert, Z.}}, \citenamefont {{Hubin, N.}}, \citenamefont {{Hummel, C. A.}}, \citenamefont {{Jakob, G.}}, \citenamefont {{Janssen, A.}}, \citenamefont {{Jochum, L.}}, \citenamefont {{Jocou,
  L.}}, \citenamefont {{Kaufer, A.}}, \citenamefont {{Kellner, S.}}, \citenamefont {{Kendrew, S.}}, \citenamefont {{Kern, L.}}, \citenamefont {{Kervella, P.}}, \citenamefont {{Kiekebusch, M.}}, \citenamefont {{Klein, R.}}, \citenamefont {{Kok, Y.}}, \citenamefont {{Kolb, J.}}, \citenamefont {{Kulas, M.}}, \citenamefont {{Lacour, S.}}, \citenamefont {{Lapeyrère, V.}}, \citenamefont {{Lazareff, B.}}, \citenamefont {{Le Bouquin, J.-B.}}, \citenamefont {{Lèna, P.}}, \citenamefont {{Lenzen, R.}}, \citenamefont {{Lévêque, S.}}, \citenamefont {{Lippa, M.}}, \citenamefont {{Magnard, Y.}}, \citenamefont {{Mehrgan, L.}}, \citenamefont {{Mellein, M.}}, \citenamefont {{Mérand, A.}}, \citenamefont {{Moreno-Ventas, J.}}, \citenamefont {{Moulin, T.}}, \citenamefont {{Müller, E.}}, \citenamefont {{Müller, F.}}, \citenamefont {{Neumann, U.}}, \citenamefont {{Oberti, S.}}, \citenamefont {{Ott, T.}}, \citenamefont {{Pallanca, L.}}, \citenamefont {{Panduro, J.}}, \citenamefont {{Pasquini, L.}}, \citenamefont {{Paumard,
  T.}}, \citenamefont {{Percheron, I.}}, \citenamefont {{Perraut, K.}}, \citenamefont {{Perrin, G.}}, \citenamefont {{Pflüger, A.}}, \citenamefont {{Pfuhl, O.}}, \citenamefont {{Phan Duc, T.}}, \citenamefont {{Plewa, P. M.}}, \citenamefont {{Popovic, D.}}, \citenamefont {{Rabien, S.}}, \citenamefont {{Ramírez, A.}}, \citenamefont {{Ramos, J.}}, \citenamefont {{Rau, C.}}, \citenamefont {{Riquelme, M.}}, \citenamefont {{Rohloff, R.-R.}}, \citenamefont {{Rousset, G.}}, \citenamefont {{Sanchez-Bermudez, J.}}, \citenamefont {{Scheithauer, S.}}, \citenamefont {{Schöller, M.}}, \citenamefont {{Schuhler, N.}}, \citenamefont {{Spyromilio, J.}}, \citenamefont {{Straubmeier, C.}}, \citenamefont {{Sturm, E.}}, \citenamefont {{Suarez, M.}}, \citenamefont {{Tristram, K. R. W.}}, \citenamefont {{Ventura, N.}}, \citenamefont {{Vincent, F.}}, \citenamefont {{Waisberg, I.}}, \citenamefont {{Wank, I.}}, \citenamefont {{Weber, J.}}, \citenamefont {{Wieprecht, E.}}, \citenamefont {{Wiest, M.}}, \citenamefont {{Wiezorrek, E.}},
  \citenamefont {{Wittkowski, M.}}, \citenamefont {{Woillez, J.}}, \citenamefont {{Wolff, B.}}, \citenamefont {{Yazici, S.}}, \citenamefont {{Ziegler, D.}},\ and\ \citenamefont {{Zins, G.}}}]{GRAVITY2017AA}%
  \BibitemOpen
  \bibfield  {author} {\bibinfo {author} {\bibnamefont {{GRAVITY Collaboration}}}, \bibinfo {author} {\bibnamefont {{Abuter, R.}}}, \bibinfo {author} {\bibnamefont {{Accardo, M.}}}, \bibinfo {author} {\bibnamefont {{Amorim, A.}}}, \bibinfo {author} {\bibnamefont {{Anugu, N.}}}, \bibinfo {author} {\bibnamefont {{Ávila, G.}}}, \bibinfo {author} {\bibnamefont {{Azouaoui, N.}}}, \bibinfo {author} {\bibnamefont {{Benisty, M.}}}, \bibinfo {author} {\bibnamefont {{Berger, J. P.}}}, \bibinfo {author} {\bibnamefont {{Blind, N.}}}, \bibinfo {author} {\bibnamefont {{Bonnet, H.}}}, \bibinfo {author} {\bibnamefont {{Bourget, P.}}}, \bibinfo {author} {\bibnamefont {{Brandner, W.}}}, \bibinfo {author} {\bibnamefont {{Brast, R.}}}, \bibinfo {author} {\bibnamefont {{Buron, A.}}}, \bibinfo {author} {\bibnamefont {{Burtscher, L.}}}, \bibinfo {author} {\bibnamefont {{Cassaing, F.}}}, \bibinfo {author} {\bibnamefont {{Chapron, F.}}}, \bibinfo {author} {\bibnamefont {{Choquet, É.}}}, \bibinfo {author} {\bibnamefont {{Clénet,
  Y.}}}, \bibinfo {author} {\bibnamefont {{Collin, C.}}}, \bibinfo {author} {\bibnamefont {{Coudé du Foresto, V.}}}, \bibinfo {author} {\bibnamefont {{de Wit, W.}}}, \bibinfo {author} {\bibnamefont {{de Zeeuw, P. T.}}}, \bibinfo {author} {\bibnamefont {{Deen, C.}}}, \bibinfo {author} {\bibnamefont {{Delplancke-Ströbele, F.}}}, \bibinfo {author} {\bibnamefont {{Dembet, R.}}}, \bibinfo {author} {\bibnamefont {{Derie, F.}}}, \bibinfo {author} {\bibnamefont {{Dexter, J.}}}, \bibinfo {author} {\bibnamefont {{Duvert, G.}}}, \bibinfo {author} {\bibnamefont {{Ebert, M.}}}, \bibinfo {author} {\bibnamefont {{Eckart, A.}}}, \bibinfo {author} {\bibnamefont {{Eisenhauer, F.}}}, \bibinfo {author} {\bibnamefont {{Esselborn, M.}}}, \bibinfo {author} {\bibnamefont {{Fédou, P.}}}, \bibinfo {author} {\bibnamefont {{Finger, G.}}}, \bibinfo {author} {\bibnamefont {{Garcia, P.}}}, \bibinfo {author} {\bibnamefont {{Garcia Dabo, C. E.}}}, \bibinfo {author} {\bibnamefont {{Garcia Lopez, R.}}}, \bibinfo {author} {\bibnamefont
  {{Gendron, E.}}}, \bibinfo {author} {\bibnamefont {{Genzel, R.}}}, \bibinfo {author} {\bibnamefont {{Gillessen, S.}}}, \bibinfo {author} {\bibnamefont {{Gonte, F.}}}, \bibinfo {author} {\bibnamefont {{Gordo, P.}}}, \bibinfo {author} {\bibnamefont {{Grould, M.}}}, \bibinfo {author} {\bibnamefont {{Grözinger, U.}}}, \bibinfo {author} {\bibnamefont {{Guieu, S.}}}, \bibinfo {author} {\bibnamefont {{Haguenauer, P.}}}, \bibinfo {author} {\bibnamefont {{Hans, O.}}}, \bibinfo {author} {\bibnamefont {{Haubois, X.}}}, \bibinfo {author} {\bibnamefont {{Haug, M.}}}, \bibinfo {author} {\bibnamefont {{Haussmann, F.}}}, \bibinfo {author} {\bibnamefont {{Henning, Th.}}}, \bibinfo {author} {\bibnamefont {{Hippler, S.}}}, \bibinfo {author} {\bibnamefont {{Horrobin, M.}}}, \bibinfo {author} {\bibnamefont {{Huber, A.}}}, \bibinfo {author} {\bibnamefont {{Hubert, Z.}}}, \bibinfo {author} {\bibnamefont {{Hubin, N.}}}, \bibinfo {author} {\bibnamefont {{Hummel, C. A.}}}, \bibinfo {author} {\bibnamefont {{Jakob, G.}}}, \bibinfo
  {author} {\bibnamefont {{Janssen, A.}}}, \bibinfo {author} {\bibnamefont {{Jochum, L.}}}, \bibinfo {author} {\bibnamefont {{Jocou, L.}}}, \bibinfo {author} {\bibnamefont {{Kaufer, A.}}}, \bibinfo {author} {\bibnamefont {{Kellner, S.}}}, \bibinfo {author} {\bibnamefont {{Kendrew, S.}}}, \bibinfo {author} {\bibnamefont {{Kern, L.}}}, \bibinfo {author} {\bibnamefont {{Kervella, P.}}}, \bibinfo {author} {\bibnamefont {{Kiekebusch, M.}}}, \bibinfo {author} {\bibnamefont {{Klein, R.}}}, \bibinfo {author} {\bibnamefont {{Kok, Y.}}}, \bibinfo {author} {\bibnamefont {{Kolb, J.}}}, \bibinfo {author} {\bibnamefont {{Kulas, M.}}}, \bibinfo {author} {\bibnamefont {{Lacour, S.}}}, \bibinfo {author} {\bibnamefont {{Lapeyrère, V.}}}, \bibinfo {author} {\bibnamefont {{Lazareff, B.}}}, \bibinfo {author} {\bibnamefont {{Le Bouquin, J.-B.}}}, \bibinfo {author} {\bibnamefont {{Lèna, P.}}}, \bibinfo {author} {\bibnamefont {{Lenzen, R.}}}, \bibinfo {author} {\bibnamefont {{Lévêque, S.}}}, \bibinfo {author} {\bibnamefont
  {{Lippa, M.}}}, \bibinfo {author} {\bibnamefont {{Magnard, Y.}}}, \bibinfo {author} {\bibnamefont {{Mehrgan, L.}}}, \bibinfo {author} {\bibnamefont {{Mellein, M.}}}, \bibinfo {author} {\bibnamefont {{Mérand, A.}}}, \bibinfo {author} {\bibnamefont {{Moreno-Ventas, J.}}}, \bibinfo {author} {\bibnamefont {{Moulin, T.}}}, \bibinfo {author} {\bibnamefont {{Müller, E.}}}, \bibinfo {author} {\bibnamefont {{Müller, F.}}}, \bibinfo {author} {\bibnamefont {{Neumann, U.}}}, \bibinfo {author} {\bibnamefont {{Oberti, S.}}}, \bibinfo {author} {\bibnamefont {{Ott, T.}}}, \bibinfo {author} {\bibnamefont {{Pallanca, L.}}}, \bibinfo {author} {\bibnamefont {{Panduro, J.}}}, \bibinfo {author} {\bibnamefont {{Pasquini, L.}}}, \bibinfo {author} {\bibnamefont {{Paumard, T.}}}, \bibinfo {author} {\bibnamefont {{Percheron, I.}}}, \bibinfo {author} {\bibnamefont {{Perraut, K.}}}, \bibinfo {author} {\bibnamefont {{Perrin, G.}}}, \bibinfo {author} {\bibnamefont {{Pflüger, A.}}}, \bibinfo {author} {\bibnamefont {{Pfuhl, O.}}},
  \bibinfo {author} {\bibnamefont {{Phan Duc, T.}}}, \bibinfo {author} {\bibnamefont {{Plewa, P. M.}}}, \bibinfo {author} {\bibnamefont {{Popovic, D.}}}, \bibinfo {author} {\bibnamefont {{Rabien, S.}}}, \bibinfo {author} {\bibnamefont {{Ramírez, A.}}}, \bibinfo {author} {\bibnamefont {{Ramos, J.}}}, \bibinfo {author} {\bibnamefont {{Rau, C.}}}, \bibinfo {author} {\bibnamefont {{Riquelme, M.}}}, \bibinfo {author} {\bibnamefont {{Rohloff, R.-R.}}}, \bibinfo {author} {\bibnamefont {{Rousset, G.}}}, \bibinfo {author} {\bibnamefont {{Sanchez-Bermudez, J.}}}, \bibinfo {author} {\bibnamefont {{Scheithauer, S.}}}, \bibinfo {author} {\bibnamefont {{Schöller, M.}}}, \bibinfo {author} {\bibnamefont {{Schuhler, N.}}}, \bibinfo {author} {\bibnamefont {{Spyromilio, J.}}}, \bibinfo {author} {\bibnamefont {{Straubmeier, C.}}}, \bibinfo {author} {\bibnamefont {{Sturm, E.}}}, \bibinfo {author} {\bibnamefont {{Suarez, M.}}}, \bibinfo {author} {\bibnamefont {{Tristram, K. R. W.}}}, \bibinfo {author} {\bibnamefont {{Ventura,
  N.}}}, \bibinfo {author} {\bibnamefont {{Vincent, F.}}}, \bibinfo {author} {\bibnamefont {{Waisberg, I.}}}, \bibinfo {author} {\bibnamefont {{Wank, I.}}}, \bibinfo {author} {\bibnamefont {{Weber, J.}}}, \bibinfo {author} {\bibnamefont {{Wieprecht, E.}}}, \bibinfo {author} {\bibnamefont {{Wiest, M.}}}, \bibinfo {author} {\bibnamefont {{Wiezorrek, E.}}}, \bibinfo {author} {\bibnamefont {{Wittkowski, M.}}}, \bibinfo {author} {\bibnamefont {{Woillez, J.}}}, \bibinfo {author} {\bibnamefont {{Wolff, B.}}}, \bibinfo {author} {\bibnamefont {{Yazici, S.}}}, \bibinfo {author} {\bibnamefont {{Ziegler, D.}}},\ and\ \bibinfo {author} {\bibnamefont {{Zins, G.}}},\ }\bibfield  {title} {\bibinfo {title} {First light for gravity: Phase referencing optical interferometry for the very large telescope interferometer},\ }\href {https://doi.org/10.1051/0004-6361/201730838} {\bibfield  {journal} {\bibinfo  {journal} {Astronomy $\&$ Asrtophysics}\ }\textbf {\bibinfo {volume} {602}},\ \bibinfo {pages} {A94} (\bibinfo {year}
  {2017})}\BibitemShut {NoStop}%
\bibitem [{\citenamefont {Colavita}\ \emph {et~al.}(2013)\citenamefont {Colavita}, \citenamefont {Wizinowich}, \citenamefont {Akeson}, \citenamefont {Ragland}, \citenamefont {Woillez}, \citenamefont {Millan-Gabet}, \citenamefont {Serabyn}, \citenamefont {Abajian}, \citenamefont {Acton}, \citenamefont {Appleby}, \citenamefont {Beletic}, \citenamefont {Beichman}, \citenamefont {Bell}, \citenamefont {Berkey}, \citenamefont {Berlin}, \citenamefont {Boden}, \citenamefont {Booth}, \citenamefont {Boutell}, \citenamefont {Chaffee}, \citenamefont {Chan}, \citenamefont {Chin}, \citenamefont {Chock}, \citenamefont {Cohen}, \citenamefont {Cooper}, \citenamefont {Crawford}, \citenamefont {Creech-Eakman}, \citenamefont {Dahl}, \citenamefont {Eychaner}, \citenamefont {Fanson}, \citenamefont {Felizardo}, \citenamefont {Garcia-Gathright}, \citenamefont {Gathright}, \citenamefont {Hardy}, \citenamefont {Henderson}, \citenamefont {Herstein}, \citenamefont {Hess}, \citenamefont {Hovland}, \citenamefont {Hrynevych}, \citenamefont
  {Johansson}, \citenamefont {Johnson}, \citenamefont {Kelley}, \citenamefont {Kendrick}, \citenamefont {Koresko}, \citenamefont {Kurpis}, \citenamefont {Mignant}, \citenamefont {Lewis}, \citenamefont {Ligon}, \citenamefont {Lupton}, \citenamefont {McBride}, \citenamefont {Medeiros}, \citenamefont {Mennesson}, \citenamefont {Moore}, \citenamefont {Morrison}, \citenamefont {Nance}, \citenamefont {Neyman}, \citenamefont {Niessner}, \citenamefont {Paine}, \citenamefont {Palmer}, \citenamefont {Panteleeva}, \citenamefont {Papin}, \citenamefont {Parvin}, \citenamefont {Reder}, \citenamefont {Rudeen}, \citenamefont {Saloga}, \citenamefont {Sargent}, \citenamefont {Shao}, \citenamefont {Smith}, \citenamefont {Smythe}, \citenamefont {Stomski}, \citenamefont {Summers}, \citenamefont {Swain}, \citenamefont {Swanson}, \citenamefont {Thompson}, \citenamefont {Tsubota}, \citenamefont {Tumminello}, \citenamefont {Tyau}, \citenamefont {van Belle}, \citenamefont {Vasisht}, \citenamefont {Vause}, \citenamefont {Vescelus},
  \citenamefont {Walker}, \citenamefont {Wallace}, \citenamefont {Wehmeier},\ and\ \citenamefont {Wetherell}}]{colavita2013PASP}%
  \BibitemOpen
  \bibfield  {author} {\bibinfo {author} {\bibfnamefont {M.~M.}\ \bibnamefont {Colavita}}, \bibinfo {author} {\bibfnamefont {P.~L.}\ \bibnamefont {Wizinowich}}, \bibinfo {author} {\bibfnamefont {R.~L.}\ \bibnamefont {Akeson}}, \bibinfo {author} {\bibfnamefont {S.}~\bibnamefont {Ragland}}, \bibinfo {author} {\bibfnamefont {J.~M.}\ \bibnamefont {Woillez}}, \bibinfo {author} {\bibfnamefont {R.}~\bibnamefont {Millan-Gabet}}, \bibinfo {author} {\bibfnamefont {E.}~\bibnamefont {Serabyn}}, \bibinfo {author} {\bibfnamefont {M.}~\bibnamefont {Abajian}}, \bibinfo {author} {\bibfnamefont {D.~S.}\ \bibnamefont {Acton}}, \bibinfo {author} {\bibfnamefont {E.}~\bibnamefont {Appleby}}, \bibinfo {author} {\bibfnamefont {J.~W.}\ \bibnamefont {Beletic}}, \bibinfo {author} {\bibfnamefont {C.~A.}\ \bibnamefont {Beichman}}, \bibinfo {author} {\bibfnamefont {J.}~\bibnamefont {Bell}}, \bibinfo {author} {\bibfnamefont {B.~C.}\ \bibnamefont {Berkey}}, \bibinfo {author} {\bibfnamefont {J.}~\bibnamefont {Berlin}}, \bibinfo {author}
  {\bibfnamefont {A.~F.}\ \bibnamefont {Boden}}, \bibinfo {author} {\bibfnamefont {A.~J.}\ \bibnamefont {Booth}}, \bibinfo {author} {\bibfnamefont {R.}~\bibnamefont {Boutell}}, \bibinfo {author} {\bibfnamefont {F.~H.}\ \bibnamefont {Chaffee}}, \bibinfo {author} {\bibfnamefont {D.}~\bibnamefont {Chan}}, \bibinfo {author} {\bibfnamefont {J.}~\bibnamefont {Chin}}, \bibinfo {author} {\bibfnamefont {J.}~\bibnamefont {Chock}}, \bibinfo {author} {\bibfnamefont {R.}~\bibnamefont {Cohen}}, \bibinfo {author} {\bibfnamefont {A.}~\bibnamefont {Cooper}}, \bibinfo {author} {\bibfnamefont {S.~L.}\ \bibnamefont {Crawford}}, \bibinfo {author} {\bibfnamefont {M.~J.}\ \bibnamefont {Creech-Eakman}}, \bibinfo {author} {\bibfnamefont {W.}~\bibnamefont {Dahl}}, \bibinfo {author} {\bibfnamefont {G.}~\bibnamefont {Eychaner}}, \bibinfo {author} {\bibfnamefont {J.~L.}\ \bibnamefont {Fanson}}, \bibinfo {author} {\bibfnamefont {C.}~\bibnamefont {Felizardo}}, \bibinfo {author} {\bibfnamefont {J.~I.}\ \bibnamefont {Garcia-Gathright}},
  \bibinfo {author} {\bibfnamefont {J.~T.}\ \bibnamefont {Gathright}}, \bibinfo {author} {\bibfnamefont {G.}~\bibnamefont {Hardy}}, \bibinfo {author} {\bibfnamefont {H.}~\bibnamefont {Henderson}}, \bibinfo {author} {\bibfnamefont {J.~S.}\ \bibnamefont {Herstein}}, \bibinfo {author} {\bibfnamefont {M.}~\bibnamefont {Hess}}, \bibinfo {author} {\bibfnamefont {E.~E.}\ \bibnamefont {Hovland}}, \bibinfo {author} {\bibfnamefont {M.~A.}\ \bibnamefont {Hrynevych}}, \bibinfo {author} {\bibfnamefont {E.}~\bibnamefont {Johansson}}, \bibinfo {author} {\bibfnamefont {R.~L.}\ \bibnamefont {Johnson}}, \bibinfo {author} {\bibfnamefont {J.}~\bibnamefont {Kelley}}, \bibinfo {author} {\bibfnamefont {R.}~\bibnamefont {Kendrick}}, \bibinfo {author} {\bibfnamefont {C.~D.}\ \bibnamefont {Koresko}}, \bibinfo {author} {\bibfnamefont {P.}~\bibnamefont {Kurpis}}, \bibinfo {author} {\bibfnamefont {D.~L.}\ \bibnamefont {Mignant}}, \bibinfo {author} {\bibfnamefont {H.~A.}\ \bibnamefont {Lewis}}, \bibinfo {author} {\bibfnamefont {E.~R.}\
  \bibnamefont {Ligon}}, \bibinfo {author} {\bibfnamefont {W.}~\bibnamefont {Lupton}}, \bibinfo {author} {\bibfnamefont {D.}~\bibnamefont {McBride}}, \bibinfo {author} {\bibfnamefont {D.~W.}\ \bibnamefont {Medeiros}}, \bibinfo {author} {\bibfnamefont {B.~P.}\ \bibnamefont {Mennesson}}, \bibinfo {author} {\bibfnamefont {J.~D.}\ \bibnamefont {Moore}}, \bibinfo {author} {\bibfnamefont {D.}~\bibnamefont {Morrison}}, \bibinfo {author} {\bibfnamefont {C.}~\bibnamefont {Nance}}, \bibinfo {author} {\bibfnamefont {C.}~\bibnamefont {Neyman}}, \bibinfo {author} {\bibfnamefont {A.}~\bibnamefont {Niessner}}, \bibinfo {author} {\bibfnamefont {C.~G.}\ \bibnamefont {Paine}}, \bibinfo {author} {\bibfnamefont {D.~L.}\ \bibnamefont {Palmer}}, \bibinfo {author} {\bibfnamefont {T.}~\bibnamefont {Panteleeva}}, \bibinfo {author} {\bibfnamefont {M.}~\bibnamefont {Papin}}, \bibinfo {author} {\bibfnamefont {B.}~\bibnamefont {Parvin}}, \bibinfo {author} {\bibfnamefont {L.}~\bibnamefont {Reder}}, \bibinfo {author} {\bibfnamefont
  {A.}~\bibnamefont {Rudeen}}, \bibinfo {author} {\bibfnamefont {T.}~\bibnamefont {Saloga}}, \bibinfo {author} {\bibfnamefont {A.}~\bibnamefont {Sargent}}, \bibinfo {author} {\bibfnamefont {M.}~\bibnamefont {Shao}}, \bibinfo {author} {\bibfnamefont {B.}~\bibnamefont {Smith}}, \bibinfo {author} {\bibfnamefont {R.~F.}\ \bibnamefont {Smythe}}, \bibinfo {author} {\bibfnamefont {P.}~\bibnamefont {Stomski}}, \bibinfo {author} {\bibfnamefont {K.~R.}\ \bibnamefont {Summers}}, \bibinfo {author} {\bibfnamefont {M.~R.}\ \bibnamefont {Swain}}, \bibinfo {author} {\bibfnamefont {P.}~\bibnamefont {Swanson}}, \bibinfo {author} {\bibfnamefont {R.}~\bibnamefont {Thompson}}, \bibinfo {author} {\bibfnamefont {K.}~\bibnamefont {Tsubota}}, \bibinfo {author} {\bibfnamefont {A.}~\bibnamefont {Tumminello}}, \bibinfo {author} {\bibfnamefont {C.}~\bibnamefont {Tyau}}, \bibinfo {author} {\bibfnamefont {G.~T.}\ \bibnamefont {van Belle}}, \bibinfo {author} {\bibfnamefont {G.}~\bibnamefont {Vasisht}}, \bibinfo {author} {\bibfnamefont
  {J.}~\bibnamefont {Vause}}, \bibinfo {author} {\bibfnamefont {F.}~\bibnamefont {Vescelus}}, \bibinfo {author} {\bibfnamefont {J.}~\bibnamefont {Walker}}, \bibinfo {author} {\bibfnamefont {J.~K.}\ \bibnamefont {Wallace}}, \bibinfo {author} {\bibfnamefont {U.}~\bibnamefont {Wehmeier}},\ and\ \bibinfo {author} {\bibfnamefont {E.}~\bibnamefont {Wetherell}},\ }\bibfield  {title} {\bibinfo {title} {The keck interferometer},\ }\href {https://doi.org/10.1086/673475} {\bibfield  {journal} {\bibinfo  {journal} {Publications of the Astronomical Society of the Pacific}\ }\textbf {\bibinfo {volume} {125}},\ \bibinfo {pages} {1226} (\bibinfo {year} {2013})}\BibitemShut {NoStop}%
\bibitem [{\citenamefont {Cody}\ \emph {et~al.}(2014)\citenamefont {Cody}, \citenamefont {Stauffer}, \citenamefont {Baglin}, \citenamefont {Micela}, \citenamefont {Rebull}, \citenamefont {Flaccomio}, \citenamefont {Morales-Calderón}, \citenamefont {Aigrain}, \citenamefont {Bouvier}, \citenamefont {Hillenbrand}, \citenamefont {Gutermuth}, \citenamefont {Song}, \citenamefont {Turner}, \citenamefont {Alencar}, \citenamefont {Zwintz}, \citenamefont {Plavchan}, \citenamefont {Carpenter}, \citenamefont {Findeisen}, \citenamefont {Carey}, \citenamefont {Terebey}, \citenamefont {Hartmann}, \citenamefont {Calvet}, \citenamefont {Teixeira}, \citenamefont {Vrba}, \citenamefont {Wolk}, \citenamefont {Covey}, \citenamefont {Poppenhaeger}, \citenamefont {Günther}, \citenamefont {Forbrich}, \citenamefont {Whitney}, \citenamefont {Affer}, \citenamefont {Herbst}, \citenamefont {Hora}, \citenamefont {Barrado}, \citenamefont {Holtzman}, \citenamefont {Marchis}, \citenamefont {Wood}, \citenamefont {Guimarães}, \citenamefont
  {Box}, \citenamefont {Gillen}, \citenamefont {McQuillan}, \citenamefont {Espaillat}, \citenamefont {Allen}, \citenamefont {D'Alessio},\ and\ \citenamefont {Favata}}]{cody2014aj}%
  \BibitemOpen
  \bibfield  {author} {\bibinfo {author} {\bibfnamefont {A.~M.}\ \bibnamefont {Cody}}, \bibinfo {author} {\bibfnamefont {J.}~\bibnamefont {Stauffer}}, \bibinfo {author} {\bibfnamefont {A.}~\bibnamefont {Baglin}}, \bibinfo {author} {\bibfnamefont {G.}~\bibnamefont {Micela}}, \bibinfo {author} {\bibfnamefont {L.~M.}\ \bibnamefont {Rebull}}, \bibinfo {author} {\bibfnamefont {E.}~\bibnamefont {Flaccomio}}, \bibinfo {author} {\bibfnamefont {M.}~\bibnamefont {Morales-Calderón}}, \bibinfo {author} {\bibfnamefont {S.}~\bibnamefont {Aigrain}}, \bibinfo {author} {\bibfnamefont {J.}~\bibnamefont {Bouvier}}, \bibinfo {author} {\bibfnamefont {L.~A.}\ \bibnamefont {Hillenbrand}}, \bibinfo {author} {\bibfnamefont {R.}~\bibnamefont {Gutermuth}}, \bibinfo {author} {\bibfnamefont {I.}~\bibnamefont {Song}}, \bibinfo {author} {\bibfnamefont {N.}~\bibnamefont {Turner}}, \bibinfo {author} {\bibfnamefont {S.~H.~P.}\ \bibnamefont {Alencar}}, \bibinfo {author} {\bibfnamefont {K.}~\bibnamefont {Zwintz}}, \bibinfo {author}
  {\bibfnamefont {P.}~\bibnamefont {Plavchan}}, \bibinfo {author} {\bibfnamefont {J.}~\bibnamefont {Carpenter}}, \bibinfo {author} {\bibfnamefont {K.}~\bibnamefont {Findeisen}}, \bibinfo {author} {\bibfnamefont {S.}~\bibnamefont {Carey}}, \bibinfo {author} {\bibfnamefont {S.}~\bibnamefont {Terebey}}, \bibinfo {author} {\bibfnamefont {L.}~\bibnamefont {Hartmann}}, \bibinfo {author} {\bibfnamefont {N.}~\bibnamefont {Calvet}}, \bibinfo {author} {\bibfnamefont {P.}~\bibnamefont {Teixeira}}, \bibinfo {author} {\bibfnamefont {F.~J.}\ \bibnamefont {Vrba}}, \bibinfo {author} {\bibfnamefont {S.}~\bibnamefont {Wolk}}, \bibinfo {author} {\bibfnamefont {K.}~\bibnamefont {Covey}}, \bibinfo {author} {\bibfnamefont {K.}~\bibnamefont {Poppenhaeger}}, \bibinfo {author} {\bibfnamefont {H.~M.}\ \bibnamefont {Günther}}, \bibinfo {author} {\bibfnamefont {J.}~\bibnamefont {Forbrich}}, \bibinfo {author} {\bibfnamefont {B.}~\bibnamefont {Whitney}}, \bibinfo {author} {\bibfnamefont {L.}~\bibnamefont {Affer}}, \bibinfo {author}
  {\bibfnamefont {W.}~\bibnamefont {Herbst}}, \bibinfo {author} {\bibfnamefont {J.}~\bibnamefont {Hora}}, \bibinfo {author} {\bibfnamefont {D.}~\bibnamefont {Barrado}}, \bibinfo {author} {\bibfnamefont {J.}~\bibnamefont {Holtzman}}, \bibinfo {author} {\bibfnamefont {F.}~\bibnamefont {Marchis}}, \bibinfo {author} {\bibfnamefont {K.}~\bibnamefont {Wood}}, \bibinfo {author} {\bibfnamefont {M.~M.}\ \bibnamefont {Guimarães}}, \bibinfo {author} {\bibfnamefont {J.~L.}\ \bibnamefont {Box}}, \bibinfo {author} {\bibfnamefont {E.}~\bibnamefont {Gillen}}, \bibinfo {author} {\bibfnamefont {A.}~\bibnamefont {McQuillan}}, \bibinfo {author} {\bibfnamefont {C.}~\bibnamefont {Espaillat}}, \bibinfo {author} {\bibfnamefont {L.}~\bibnamefont {Allen}}, \bibinfo {author} {\bibfnamefont {P.}~\bibnamefont {D'Alessio}},\ and\ \bibinfo {author} {\bibfnamefont {F.}~\bibnamefont {Favata}},\ }\bibfield  {title} {\bibinfo {title} {Csi 2264: Simultaneous optical and infrared light curves of young disk-bearing stars in ngc-2264 with corot
  and spitzer-evidence for multiple origins of variability*},\ }\href {https://doi.org/10.1088/0004-6256/147/4/82} {\bibfield  {journal} {\bibinfo  {journal} {The Astronomical Journal}\ }\textbf {\bibinfo {volume} {147}},\ \bibinfo {pages} {82} (\bibinfo {year} {2014})}\BibitemShut {NoStop}%
\bibitem [{\citenamefont {Marigo}\ \emph {et~al.}(2017)\citenamefont {Marigo}, \citenamefont {Girardi}, \citenamefont {Bressan}, \citenamefont {Rosenfield}, \citenamefont {Aringer}, \citenamefont {Chen}, \citenamefont {Dussin}, \citenamefont {Nanni}, \citenamefont {Pastorelli}, \citenamefont {Rodrigues}, \citenamefont {Trabucchi}, \citenamefont {Bladh}, \citenamefont {Dalcanton}, \citenamefont {Groenewegen}, \citenamefont {Montalbán},\ and\ \citenamefont {Wood}}]{marigo2017aj}%
  \BibitemOpen
  \bibfield  {author} {\bibinfo {author} {\bibfnamefont {P.}~\bibnamefont {Marigo}}, \bibinfo {author} {\bibfnamefont {L.}~\bibnamefont {Girardi}}, \bibinfo {author} {\bibfnamefont {A.}~\bibnamefont {Bressan}}, \bibinfo {author} {\bibfnamefont {P.}~\bibnamefont {Rosenfield}}, \bibinfo {author} {\bibfnamefont {B.}~\bibnamefont {Aringer}}, \bibinfo {author} {\bibfnamefont {Y.}~\bibnamefont {Chen}}, \bibinfo {author} {\bibfnamefont {M.}~\bibnamefont {Dussin}}, \bibinfo {author} {\bibfnamefont {A.}~\bibnamefont {Nanni}}, \bibinfo {author} {\bibfnamefont {G.}~\bibnamefont {Pastorelli}}, \bibinfo {author} {\bibfnamefont {T.~S.}\ \bibnamefont {Rodrigues}}, \bibinfo {author} {\bibfnamefont {M.}~\bibnamefont {Trabucchi}}, \bibinfo {author} {\bibfnamefont {S.}~\bibnamefont {Bladh}}, \bibinfo {author} {\bibfnamefont {J.}~\bibnamefont {Dalcanton}}, \bibinfo {author} {\bibfnamefont {M.~A.~T.}\ \bibnamefont {Groenewegen}}, \bibinfo {author} {\bibfnamefont {J.}~\bibnamefont {Montalbán}},\ and\ \bibinfo {author}
  {\bibfnamefont {P.~R.}\ \bibnamefont {Wood}},\ }\bibfield  {title} {\bibinfo {title} {A new generation of parsec-colibri stellar isochrones including the tp-agb phase},\ }\href {https://doi.org/10.3847/1538-4357/835/1/77} {\bibfield  {journal} {\bibinfo  {journal} {The Astrophysical Journal}\ }\textbf {\bibinfo {volume} {835}},\ \bibinfo {pages} {77} (\bibinfo {year} {2017})}\BibitemShut {NoStop}%
\bibitem [{\citenamefont {Wang}\ and\ \citenamefont {Chen}(2019)}]{wang2019aj}%
  \BibitemOpen
  \bibfield  {author} {\bibinfo {author} {\bibfnamefont {S.}~\bibnamefont {Wang}}\ and\ \bibinfo {author} {\bibfnamefont {X.}~\bibnamefont {Chen}},\ }\bibfield  {title} {\bibinfo {title} {The optical to mid-infrared extinction law based on the apogee, gaia dr2, pan-starrs1, sdss, apass, 2mass, and wise surveys},\ }\href {https://doi.org/10.3847/1538-4357/ab1c61} {\bibfield  {journal} {\bibinfo  {journal} {The Astrophysical Journal}\ }\textbf {\bibinfo {volume} {877}},\ \bibinfo {pages} {116} (\bibinfo {year} {2019})}\BibitemShut {NoStop}%
\bibitem [{\citenamefont {Quanz}\ \emph {et~al.}(2022)\citenamefont {Quanz}, \citenamefont {Ottiger}, \citenamefont {Fontanet}, \citenamefont {Kammerer}, \citenamefont {Menti}, \citenamefont {Dannert}, \citenamefont {Gheorghe}, \citenamefont {Absil}, \citenamefont {Airapetian}, \citenamefont {Alei}, \citenamefont {Allart}, \citenamefont {Angerhausen}, \citenamefont {Blumenthal}, \citenamefont {Buchhave}, \citenamefont {Cabrera}, \citenamefont {Carrión-González}, \citenamefont {Chauvin}, \citenamefont {Danchi}, \citenamefont {Dandumont}, \citenamefont {Defrére}, \citenamefont {Dorn}, \citenamefont {Ehrenreich}, \citenamefont {Ertel}, \citenamefont {Fridlund}, \citenamefont {García~Muñoz}, \citenamefont {Gascón}, \citenamefont {Girard}, \citenamefont {Glauser}, \citenamefont {Grenfell}, \citenamefont {Guidi}, \citenamefont {Hagelberg}, \citenamefont {Helled}, \citenamefont {Ireland}, \citenamefont {Janson}, \citenamefont {Kopparapu}, \citenamefont {Korth}, \citenamefont {Kozakis}, \citenamefont {Kraus},
  \citenamefont {Léger}, \citenamefont {Leedjärv}, \citenamefont {Lichtenberg}, \citenamefont {Lillo-Box}, \citenamefont {Linz}, \citenamefont {Liseau}, \citenamefont {Loicq}, \citenamefont {Mahendra}, \citenamefont {Malbet}, \citenamefont {Mathew}, \citenamefont {Mennesson}, \citenamefont {Meyer}, \citenamefont {Mishra}, \citenamefont {Molaverdikhani}, \citenamefont {Noack}, \citenamefont {Oza}, \citenamefont {Pallé}, \citenamefont {Parviainen}, \citenamefont {Quirrenbach}, \citenamefont {Rauer}, \citenamefont {Ribas}, \citenamefont {Rice}, \citenamefont {Romagnolo}, \citenamefont {Rugheimer}, \citenamefont {Schwieterman}, \citenamefont {Serabyn}, \citenamefont {Sharma}, \citenamefont {Stassun}, \citenamefont {Szulágyi}, \citenamefont {Wang}, \citenamefont {Wunderlich}, \citenamefont {Wyatt},\ and\ \citenamefont {Collaboration}}]{quanz2022aa}%
  \BibitemOpen
  \bibfield  {author} {\bibinfo {author} {\bibfnamefont {S.~P.}\ \bibnamefont {Quanz}}, \bibinfo {author} {\bibfnamefont {M.}~\bibnamefont {Ottiger}}, \bibinfo {author} {\bibfnamefont {E.}~\bibnamefont {Fontanet}}, \bibinfo {author} {\bibfnamefont {J.}~\bibnamefont {Kammerer}}, \bibinfo {author} {\bibfnamefont {F.}~\bibnamefont {Menti}}, \bibinfo {author} {\bibfnamefont {F.}~\bibnamefont {Dannert}}, \bibinfo {author} {\bibfnamefont {A.}~\bibnamefont {Gheorghe}}, \bibinfo {author} {\bibfnamefont {O.}~\bibnamefont {Absil}}, \bibinfo {author} {\bibfnamefont {V.~S.}\ \bibnamefont {Airapetian}}, \bibinfo {author} {\bibfnamefont {E.}~\bibnamefont {Alei}}, \bibinfo {author} {\bibfnamefont {R.}~\bibnamefont {Allart}}, \bibinfo {author} {\bibfnamefont {D.}~\bibnamefont {Angerhausen}}, \bibinfo {author} {\bibfnamefont {S.}~\bibnamefont {Blumenthal}}, \bibinfo {author} {\bibfnamefont {L.~A.}\ \bibnamefont {Buchhave}}, \bibinfo {author} {\bibfnamefont {J.}~\bibnamefont {Cabrera}}, \bibinfo {author} {\bibfnamefont
  {Ã.}~\bibnamefont {Carrión-González}}, \bibinfo {author} {\bibfnamefont {G.}~\bibnamefont {Chauvin}}, \bibinfo {author} {\bibfnamefont {W.~C.}\ \bibnamefont {Danchi}}, \bibinfo {author} {\bibfnamefont {C.}~\bibnamefont {Dandumont}}, \bibinfo {author} {\bibfnamefont {D.}~\bibnamefont {Defrére}}, \bibinfo {author} {\bibfnamefont {C.}~\bibnamefont {Dorn}}, \bibinfo {author} {\bibfnamefont {D.}~\bibnamefont {Ehrenreich}}, \bibinfo {author} {\bibfnamefont {S.}~\bibnamefont {Ertel}}, \bibinfo {author} {\bibfnamefont {M.}~\bibnamefont {Fridlund}}, \bibinfo {author} {\bibfnamefont {A.}~\bibnamefont {García~Muñoz}}, \bibinfo {author} {\bibfnamefont {C.}~\bibnamefont {Gascón}}, \bibinfo {author} {\bibfnamefont {J.~H.}\ \bibnamefont {Girard}}, \bibinfo {author} {\bibfnamefont {A.}~\bibnamefont {Glauser}}, \bibinfo {author} {\bibfnamefont {J.~L.}\ \bibnamefont {Grenfell}}, \bibinfo {author} {\bibfnamefont {G.}~\bibnamefont {Guidi}}, \bibinfo {author} {\bibfnamefont {J.}~\bibnamefont {Hagelberg}}, \bibinfo
  {author} {\bibfnamefont {R.}~\bibnamefont {Helled}}, \bibinfo {author} {\bibfnamefont {M.~J.}\ \bibnamefont {Ireland}}, \bibinfo {author} {\bibfnamefont {M.}~\bibnamefont {Janson}}, \bibinfo {author} {\bibfnamefont {R.~K.}\ \bibnamefont {Kopparapu}}, \bibinfo {author} {\bibfnamefont {J.}~\bibnamefont {Korth}}, \bibinfo {author} {\bibfnamefont {T.}~\bibnamefont {Kozakis}}, \bibinfo {author} {\bibfnamefont {S.}~\bibnamefont {Kraus}}, \bibinfo {author} {\bibfnamefont {A.}~\bibnamefont {Léger}}, \bibinfo {author} {\bibfnamefont {L.}~\bibnamefont {Leedjärv}}, \bibinfo {author} {\bibfnamefont {T.}~\bibnamefont {Lichtenberg}}, \bibinfo {author} {\bibfnamefont {J.}~\bibnamefont {Lillo-Box}}, \bibinfo {author} {\bibfnamefont {H.}~\bibnamefont {Linz}}, \bibinfo {author} {\bibfnamefont {R.}~\bibnamefont {Liseau}}, \bibinfo {author} {\bibfnamefont {J.}~\bibnamefont {Loicq}}, \bibinfo {author} {\bibfnamefont {V.}~\bibnamefont {Mahendra}}, \bibinfo {author} {\bibfnamefont {F.}~\bibnamefont {Malbet}}, \bibinfo {author}
  {\bibfnamefont {J.}~\bibnamefont {Mathew}}, \bibinfo {author} {\bibfnamefont {B.}~\bibnamefont {Mennesson}}, \bibinfo {author} {\bibfnamefont {M.~R.}\ \bibnamefont {Meyer}}, \bibinfo {author} {\bibfnamefont {L.}~\bibnamefont {Mishra}}, \bibinfo {author} {\bibfnamefont {K.}~\bibnamefont {Molaverdikhani}}, \bibinfo {author} {\bibfnamefont {L.}~\bibnamefont {Noack}}, \bibinfo {author} {\bibfnamefont {A.~V.}\ \bibnamefont {Oza}}, \bibinfo {author} {\bibfnamefont {E.}~\bibnamefont {Pallé}}, \bibinfo {author} {\bibfnamefont {H.}~\bibnamefont {Parviainen}}, \bibinfo {author} {\bibfnamefont {A.}~\bibnamefont {Quirrenbach}}, \bibinfo {author} {\bibfnamefont {H.}~\bibnamefont {Rauer}}, \bibinfo {author} {\bibfnamefont {I.}~\bibnamefont {Ribas}}, \bibinfo {author} {\bibfnamefont {M.}~\bibnamefont {Rice}}, \bibinfo {author} {\bibfnamefont {A.}~\bibnamefont {Romagnolo}}, \bibinfo {author} {\bibfnamefont {S.}~\bibnamefont {Rugheimer}}, \bibinfo {author} {\bibfnamefont {E.~W.}\ \bibnamefont {Schwieterman}}, \bibinfo
  {author} {\bibfnamefont {E.}~\bibnamefont {Serabyn}}, \bibinfo {author} {\bibfnamefont {S.}~\bibnamefont {Sharma}}, \bibinfo {author} {\bibfnamefont {K.~G.}\ \bibnamefont {Stassun}}, \bibinfo {author} {\bibfnamefont {J.}~\bibnamefont {Szulágyi}}, \bibinfo {author} {\bibfnamefont {H.~S.}\ \bibnamefont {Wang}}, \bibinfo {author} {\bibfnamefont {F.}~\bibnamefont {Wunderlich}}, \bibinfo {author} {\bibfnamefont {M.~C.}\ \bibnamefont {Wyatt}},\ and\ \bibinfo {author} {\bibfnamefont {t.~L.}\ \bibnamefont {Collaboration}},\ }\bibfield  {title} {\bibinfo {title} {Large interferometer for exoplanets (life)},\ }\href {https://doi.org/10.1051/0004-6361/202140366} {\bibfield  {journal} {\bibinfo  {journal} {Astronomy $\&$ Astrophysics}\ }\textbf {\bibinfo {volume} {664}},\ \bibinfo {pages} {A21} (\bibinfo {year} {2022})}\BibitemShut {NoStop}%
\bibitem [{\citenamefont {Thilker}\ \emph {et~al.}(2023)\citenamefont {Thilker}, \citenamefont {Lee}, \citenamefont {Deger}, \citenamefont {Barnes}, \citenamefont {Bigiel}, \citenamefont {Boquien}, \citenamefont {Cao}, \citenamefont {Chevance}, \citenamefont {Dale}, \citenamefont {Egorov}, \citenamefont {Glover}, \citenamefont {Grasha}, \citenamefont {Henshaw}, \citenamefont {Klessen}, \citenamefont {Koch}, \citenamefont {Kruijssen}, \citenamefont {Leroy}, \citenamefont {Lessing}, \citenamefont {Meidt}, \citenamefont {Pinna}, \citenamefont {Querejeta}, \citenamefont {Rosolowsky}, \citenamefont {Sandstrom}, \citenamefont {Schinnerer}, \citenamefont {Smith}, \citenamefont {Watkins}, \citenamefont {Williams}, \citenamefont {Anand}, \citenamefont {Belfiore}, \citenamefont {Blanc}, \citenamefont {Chandar}, \citenamefont {Congiu}, \citenamefont {Emsellem}, \citenamefont {Groves}, \citenamefont {Kreckel}, \citenamefont {Larson}, \citenamefont {Liu}, \citenamefont {Pessa},\ and\ \citenamefont
  {Whitmore}}]{thilker2023ajl}%
  \BibitemOpen
  \bibfield  {author} {\bibinfo {author} {\bibfnamefont {D.~A.}\ \bibnamefont {Thilker}}, \bibinfo {author} {\bibfnamefont {J.~C.}\ \bibnamefont {Lee}}, \bibinfo {author} {\bibfnamefont {S.}~\bibnamefont {Deger}}, \bibinfo {author} {\bibfnamefont {A.~T.}\ \bibnamefont {Barnes}}, \bibinfo {author} {\bibfnamefont {F.}~\bibnamefont {Bigiel}}, \bibinfo {author} {\bibfnamefont {M.}~\bibnamefont {Boquien}}, \bibinfo {author} {\bibfnamefont {Y.}~\bibnamefont {Cao}}, \bibinfo {author} {\bibfnamefont {M.}~\bibnamefont {Chevance}}, \bibinfo {author} {\bibfnamefont {D.~A.}\ \bibnamefont {Dale}}, \bibinfo {author} {\bibfnamefont {O.~V.}\ \bibnamefont {Egorov}}, \bibinfo {author} {\bibfnamefont {S.~C.~O.}\ \bibnamefont {Glover}}, \bibinfo {author} {\bibfnamefont {K.}~\bibnamefont {Grasha}}, \bibinfo {author} {\bibfnamefont {J.~D.}\ \bibnamefont {Henshaw}}, \bibinfo {author} {\bibfnamefont {R.~S.}\ \bibnamefont {Klessen}}, \bibinfo {author} {\bibfnamefont {E.}~\bibnamefont {Koch}}, \bibinfo {author} {\bibfnamefont
  {J.~M.~D.}\ \bibnamefont {Kruijssen}}, \bibinfo {author} {\bibfnamefont {A.~K.}\ \bibnamefont {Leroy}}, \bibinfo {author} {\bibfnamefont {R.~A.}\ \bibnamefont {Lessing}}, \bibinfo {author} {\bibfnamefont {S.~E.}\ \bibnamefont {Meidt}}, \bibinfo {author} {\bibfnamefont {F.}~\bibnamefont {Pinna}}, \bibinfo {author} {\bibfnamefont {M.}~\bibnamefont {Querejeta}}, \bibinfo {author} {\bibfnamefont {E.}~\bibnamefont {Rosolowsky}}, \bibinfo {author} {\bibfnamefont {K.~M.}\ \bibnamefont {Sandstrom}}, \bibinfo {author} {\bibfnamefont {E.}~\bibnamefont {Schinnerer}}, \bibinfo {author} {\bibfnamefont {R.~J.}\ \bibnamefont {Smith}}, \bibinfo {author} {\bibfnamefont {E.~J.}\ \bibnamefont {Watkins}}, \bibinfo {author} {\bibfnamefont {T.~G.}\ \bibnamefont {Williams}}, \bibinfo {author} {\bibfnamefont {G.~S.}\ \bibnamefont {Anand}}, \bibinfo {author} {\bibfnamefont {F.}~\bibnamefont {Belfiore}}, \bibinfo {author} {\bibfnamefont {G.~A.}\ \bibnamefont {Blanc}}, \bibinfo {author} {\bibfnamefont {R.}~\bibnamefont {Chandar}},
  \bibinfo {author} {\bibfnamefont {E.}~\bibnamefont {Congiu}}, \bibinfo {author} {\bibfnamefont {E.}~\bibnamefont {Emsellem}}, \bibinfo {author} {\bibfnamefont {B.}~\bibnamefont {Groves}}, \bibinfo {author} {\bibfnamefont {K.}~\bibnamefont {Kreckel}}, \bibinfo {author} {\bibfnamefont {K.~L.}\ \bibnamefont {Larson}}, \bibinfo {author} {\bibfnamefont {D.}~\bibnamefont {Liu}}, \bibinfo {author} {\bibfnamefont {I.}~\bibnamefont {Pessa}},\ and\ \bibinfo {author} {\bibfnamefont {B.~C.}\ \bibnamefont {Whitmore}},\ }\bibfield  {title} {\bibinfo {title} {Phangs–jwst first results: The dust filament network of ngc 628 and its relation to star formation activity},\ }\href {https://doi.org/10.3847/2041-8213/acaeac} {\bibfield  {journal} {\bibinfo  {journal} {The Astrophysical Journal Letters}\ }\textbf {\bibinfo {volume} {944}},\ \bibinfo {pages} {L13} (\bibinfo {year} {2023})}\BibitemShut {NoStop}%
\bibitem [{\citenamefont {Han}\ \emph {et~al.}(2023)\citenamefont {Han}, \citenamefont {Zhou},\ and\ \citenamefont {Shi}}]{han2023adi}%
  \BibitemOpen
  \bibfield  {author} {\bibinfo {author} {\bibfnamefont {Z.-Q.-Z.}\ \bibnamefont {Han}}, \bibinfo {author} {\bibfnamefont {Z.-Y.}\ \bibnamefont {Zhou}},\ and\ \bibinfo {author} {\bibfnamefont {B.-S.}\ \bibnamefont {Shi}},\ }\bibfield  {title} {\bibinfo {title} {Quantum frequency transducer and its applications},\ }\href {https://doi.org/doi:10.34133/adi.0030} {\bibfield  {journal} {\bibinfo  {journal} {Advanced Devices $\&$ Instrumentation}\ }\textbf {\bibinfo {volume} {4}},\ \bibinfo {pages} {0030} (\bibinfo {year} {2023})}\BibitemShut {NoStop}%
\bibitem [{\citenamefont {Mancinelli}\ \emph {et~al.}(2017)\citenamefont {Mancinelli}, \citenamefont {Trenti}, \citenamefont {Piccione}, \citenamefont {Fontana}, \citenamefont {Dam}, \citenamefont {Tidemand-Lichtenberg}, \citenamefont {Pedersen},\ and\ \citenamefont {Pavesi}}]{mancinelli2017nc}%
  \BibitemOpen
  \bibfield  {author} {\bibinfo {author} {\bibfnamefont {M.}~\bibnamefont {Mancinelli}}, \bibinfo {author} {\bibfnamefont {A.}~\bibnamefont {Trenti}}, \bibinfo {author} {\bibfnamefont {S.}~\bibnamefont {Piccione}}, \bibinfo {author} {\bibfnamefont {G.}~\bibnamefont {Fontana}}, \bibinfo {author} {\bibfnamefont {J.~S.}\ \bibnamefont {Dam}}, \bibinfo {author} {\bibfnamefont {P.}~\bibnamefont {Tidemand-Lichtenberg}}, \bibinfo {author} {\bibfnamefont {C.}~\bibnamefont {Pedersen}},\ and\ \bibinfo {author} {\bibfnamefont {L.}~\bibnamefont {Pavesi}},\ }\bibfield  {title} {\bibinfo {title} {Mid-infrared coincidence measurements on twin photons at room temperature},\ }\href {https://doi.org/10.1038/ncomms15184} {\bibfield  {journal} {\bibinfo  {journal} {Nature Communications}\ }\textbf {\bibinfo {volume} {8}},\ \bibinfo {pages} {15184} (\bibinfo {year} {2017})}\BibitemShut {NoStop}%
\bibitem [{\citenamefont {Ge}\ \emph {et~al.}(2022)\citenamefont {Ge}, \citenamefont {Yang}, \citenamefont {Li}, \citenamefont {Li}, \citenamefont {Liu}, \citenamefont {Niu}, \citenamefont {Zhou},\ and\ \citenamefont {Shi}}]{ge2022cpb}%
  \BibitemOpen
  \bibfield  {author} {\bibinfo {author} {\bibfnamefont {Z.}~\bibnamefont {Ge}}, \bibinfo {author} {\bibfnamefont {C.}~\bibnamefont {Yang}}, \bibinfo {author} {\bibfnamefont {Y.-H.}\ \bibnamefont {Li}}, \bibinfo {author} {\bibfnamefont {Y.}~\bibnamefont {Li}}, \bibinfo {author} {\bibfnamefont {S.-K.}\ \bibnamefont {Liu}}, \bibinfo {author} {\bibfnamefont {S.-J.}\ \bibnamefont {Niu}}, \bibinfo {author} {\bibfnamefont {Z.-Y.}\ \bibnamefont {Zhou}},\ and\ \bibinfo {author} {\bibfnamefont {B.-S.}\ \bibnamefont {Shi}},\ }\bibfield  {title} {\bibinfo {title} {Up-conversion detection of mid-infrared light carrying orbital angular momentum},\ }\href {https://doi.org/10.1088/1674-1056/ac6eda} {\bibfield  {journal} {\bibinfo  {journal} {Chinese Physics B}\ }\textbf {\bibinfo {volume} {31}},\ \bibinfo {pages} {104210} (\bibinfo {year} {2022})}\BibitemShut {NoStop}%
\bibitem [{\citenamefont {Ge}\ \emph {et~al.}(2023{\natexlab{a}})\citenamefont {Ge}, \citenamefont {Zhou}, \citenamefont {Ceng}, \citenamefont {Chen}, \citenamefont {Li}, \citenamefont {Li}, \citenamefont {Niu},\ and\ \citenamefont {Shi}}]{ge2023apn}%
  \BibitemOpen
  \bibfield  {author} {\bibinfo {author} {\bibfnamefont {Z.}~\bibnamefont {Ge}}, \bibinfo {author} {\bibfnamefont {Z.-Y.}\ \bibnamefont {Zhou}}, \bibinfo {author} {\bibfnamefont {J.-X.}\ \bibnamefont {Ceng}}, \bibinfo {author} {\bibfnamefont {L.}~\bibnamefont {Chen}}, \bibinfo {author} {\bibfnamefont {Y.-H.}\ \bibnamefont {Li}}, \bibinfo {author} {\bibfnamefont {Y.}~\bibnamefont {Li}}, \bibinfo {author} {\bibfnamefont {S.-J.}\ \bibnamefont {Niu}},\ and\ \bibinfo {author} {\bibfnamefont {B.-S.}\ \bibnamefont {Shi}},\ }\bibfield  {title} {\bibinfo {title} {Thermal camera based on frequency upconversion and its noise-equivalent temperature difference characterization},\ }\href {https://doi.org/10.1117/1.Apn.2.4.046002} {\bibfield  {journal} {\bibinfo  {journal} {Advanced Photonics Nexus}\ }\textbf {\bibinfo {volume} {2}},\ \bibinfo {pages} {046002} (\bibinfo {year} {2023}{\natexlab{a}})}\BibitemShut {NoStop}%
\bibitem [{\citenamefont {Ge}\ \emph {et~al.}(2023{\natexlab{b}})\citenamefont {Ge}, \citenamefont {Han}, \citenamefont {Liu}, \citenamefont {Wang}, \citenamefont {Zhou}, \citenamefont {Yang}, \citenamefont {Li}, \citenamefont {Li}, \citenamefont {Chen}, \citenamefont {Li}, \citenamefont {Niu},\ and\ \citenamefont {Shi}}]{ge2023prapplied}%
  \BibitemOpen
  \bibfield  {author} {\bibinfo {author} {\bibfnamefont {Z.}~\bibnamefont {Ge}}, \bibinfo {author} {\bibfnamefont {Z.-Q.-Z.}\ \bibnamefont {Han}}, \bibinfo {author} {\bibfnamefont {Y.-Y.}\ \bibnamefont {Liu}}, \bibinfo {author} {\bibfnamefont {X.-H.}\ \bibnamefont {Wang}}, \bibinfo {author} {\bibfnamefont {Z.-Y.}\ \bibnamefont {Zhou}}, \bibinfo {author} {\bibfnamefont {F.}~\bibnamefont {Yang}}, \bibinfo {author} {\bibfnamefont {Y.-H.}\ \bibnamefont {Li}}, \bibinfo {author} {\bibfnamefont {Y.}~\bibnamefont {Li}}, \bibinfo {author} {\bibfnamefont {L.}~\bibnamefont {Chen}}, \bibinfo {author} {\bibfnamefont {W.-Z.}\ \bibnamefont {Li}}, \bibinfo {author} {\bibfnamefont {S.-J.}\ \bibnamefont {Niu}},\ and\ \bibinfo {author} {\bibfnamefont {B.-S.}\ \bibnamefont {Shi}},\ }\bibfield  {title} {\bibinfo {title} {Midinfrared up-conversion imaging under different illumination conditions},\ }\href {https://doi.org/10.1103/PhysRevApplied.20.054060} {\bibfield  {journal} {\bibinfo  {journal} {Physical Review Applied}\
  }\textbf {\bibinfo {volume} {20}},\ \bibinfo {pages} {054060} (\bibinfo {year} {2023}{\natexlab{b}})}\BibitemShut {NoStop}%
\bibitem [{\citenamefont {Dam}\ \emph {et~al.}(2012)\citenamefont {Dam}, \citenamefont {Tidemand-Lichtenberg},\ and\ \citenamefont {Pedersen}}]{dam2012np}%
  \BibitemOpen
  \bibfield  {author} {\bibinfo {author} {\bibfnamefont {J.~S.}\ \bibnamefont {Dam}}, \bibinfo {author} {\bibfnamefont {P.}~\bibnamefont {Tidemand-Lichtenberg}},\ and\ \bibinfo {author} {\bibfnamefont {C.}~\bibnamefont {Pedersen}},\ }\bibfield  {title} {\bibinfo {title} {Room-temperature mid-infrared single-photon spectral imaging},\ }\href {https://doi.org/10.1038/nphoton.2012.231} {\bibfield  {journal} {\bibinfo  {journal} {Nature Photonics}\ }\textbf {\bibinfo {volume} {6}},\ \bibinfo {pages} {788} (\bibinfo {year} {2012})}\BibitemShut {NoStop}%
\bibitem [{\citenamefont {Ceus}\ \emph {et~al.}(2012)\citenamefont {Ceus}, \citenamefont {Reynaud}, \citenamefont {Woillez}, \citenamefont {Lai}, \citenamefont {Delage}, \citenamefont {Grossard}, \citenamefont {Baudoin}, \citenamefont {Gomes}, \citenamefont {Bouyeron}, \citenamefont {Herrmann},\ and\ \citenamefont {Sohler}}]{ceus2012mnrasl}%
  \BibitemOpen
  \bibfield  {author} {\bibinfo {author} {\bibfnamefont {D.}~\bibnamefont {Ceus}}, \bibinfo {author} {\bibfnamefont {F.}~\bibnamefont {Reynaud}}, \bibinfo {author} {\bibfnamefont {J.}~\bibnamefont {Woillez}}, \bibinfo {author} {\bibfnamefont {O.}~\bibnamefont {Lai}}, \bibinfo {author} {\bibfnamefont {L.}~\bibnamefont {Delage}}, \bibinfo {author} {\bibfnamefont {L.}~\bibnamefont {Grossard}}, \bibinfo {author} {\bibfnamefont {R.}~\bibnamefont {Baudoin}}, \bibinfo {author} {\bibfnamefont {J.-T.}\ \bibnamefont {Gomes}}, \bibinfo {author} {\bibfnamefont {L.}~\bibnamefont {Bouyeron}}, \bibinfo {author} {\bibfnamefont {H.}~\bibnamefont {Herrmann}},\ and\ \bibinfo {author} {\bibfnamefont {W.}~\bibnamefont {Sohler}},\ }\bibfield  {title} {\bibinfo {title} {Application of frequency conversion of starlight to high-resolution imaging interferometry. on-sky sensitivity test of a single arm of the interferometer},\ }\href {https://doi.org/10.1111/j.1745-3933.2012.01352.x} {\bibfield  {journal} {\bibinfo  {journal} {Monthly
  Notices of the Royal Astronomical Society: Letters}\ }\textbf {\bibinfo {volume} {427}},\ \bibinfo {pages} {L95} (\bibinfo {year} {2012})}\BibitemShut {NoStop}%
\bibitem [{\citenamefont {Gomes}\ \emph {et~al.}(2014)\citenamefont {Gomes}, \citenamefont {Delage}, \citenamefont {Baudoin}, \citenamefont {Grossard}, \citenamefont {Bouyeron}, \citenamefont {Ceus}, \citenamefont {Reynaud}, \citenamefont {Herrmann},\ and\ \citenamefont {Sohler}}]{gomes2014prl}%
  \BibitemOpen
  \bibfield  {author} {\bibinfo {author} {\bibfnamefont {J.~T.}\ \bibnamefont {Gomes}}, \bibinfo {author} {\bibfnamefont {L.}~\bibnamefont {Delage}}, \bibinfo {author} {\bibfnamefont {R.}~\bibnamefont {Baudoin}}, \bibinfo {author} {\bibfnamefont {L.}~\bibnamefont {Grossard}}, \bibinfo {author} {\bibfnamefont {L.}~\bibnamefont {Bouyeron}}, \bibinfo {author} {\bibfnamefont {D.}~\bibnamefont {Ceus}}, \bibinfo {author} {\bibfnamefont {F.}~\bibnamefont {Reynaud}}, \bibinfo {author} {\bibfnamefont {H.}~\bibnamefont {Herrmann}},\ and\ \bibinfo {author} {\bibfnamefont {W.}~\bibnamefont {Sohler}},\ }\bibfield  {title} {\bibinfo {title} {Laboratory demonstration of spatial-coherence analysis of a blackbody through an up-conversion interferometer},\ }\href {https://doi.org/10.1103/PhysRevLett.112.143904} {\bibfield  {journal} {\bibinfo  {journal} {Physical Review Letters}\ }\textbf {\bibinfo {volume} {112}},\ \bibinfo {pages} {143904} (\bibinfo {year} {2014})}\BibitemShut {NoStop}%
\bibitem [{\citenamefont {Darr\'e}\ \emph {et~al.}(2016)\citenamefont {Darr\'e}, \citenamefont {Baudoin}, \citenamefont {Gomes}, \citenamefont {Scott}, \citenamefont {Delage}, \citenamefont {Grossard}, \citenamefont {Sturmann}, \citenamefont {Farrington}, \citenamefont {Reynaud},\ and\ \citenamefont {Brummelaar}}]{darre2016pr}%
  \BibitemOpen
  \bibfield  {author} {\bibinfo {author} {\bibfnamefont {P.}~\bibnamefont {Darr\'e}}, \bibinfo {author} {\bibfnamefont {R.}~\bibnamefont {Baudoin}}, \bibinfo {author} {\bibfnamefont {J.-T.}\ \bibnamefont {Gomes}}, \bibinfo {author} {\bibfnamefont {N.~J.}\ \bibnamefont {Scott}}, \bibinfo {author} {\bibfnamefont {L.}~\bibnamefont {Delage}}, \bibinfo {author} {\bibfnamefont {L.}~\bibnamefont {Grossard}}, \bibinfo {author} {\bibfnamefont {J.}~\bibnamefont {Sturmann}}, \bibinfo {author} {\bibfnamefont {C.}~\bibnamefont {Farrington}}, \bibinfo {author} {\bibfnamefont {F.}~\bibnamefont {Reynaud}},\ and\ \bibinfo {author} {\bibfnamefont {T.~A.~T.}\ \bibnamefont {Brummelaar}},\ }\bibfield  {title} {\bibinfo {title} {First on-sky fringes with an up-conversion interferometer tested on a telescope array},\ }\href {https://doi.org/10.1103/PhysRevLett.117.233902} {\bibfield  {journal} {\bibinfo  {journal} {Phys. Rev. Lett.}\ }\textbf {\bibinfo {volume} {117}},\ \bibinfo {pages} {233902} (\bibinfo {year}
  {2016})}\BibitemShut {NoStop}%
\bibitem [{\citenamefont {Lehmann}\ \emph {et~al.}(2018)\citenamefont {Lehmann}, \citenamefont {Darré}, \citenamefont {Boulogne}, \citenamefont {Delage}, \citenamefont {Grossard},\ and\ \citenamefont {Reynaud}}]{lehmann2018mnras}%
  \BibitemOpen
  \bibfield  {author} {\bibinfo {author} {\bibfnamefont {L.}~\bibnamefont {Lehmann}}, \bibinfo {author} {\bibfnamefont {P.}~\bibnamefont {Darré}}, \bibinfo {author} {\bibfnamefont {H.}~\bibnamefont {Boulogne}}, \bibinfo {author} {\bibfnamefont {L.}~\bibnamefont {Delage}}, \bibinfo {author} {\bibfnamefont {L.}~\bibnamefont {Grossard}},\ and\ \bibinfo {author} {\bibfnamefont {F.}~\bibnamefont {Reynaud}},\ }\bibfield  {title} {\bibinfo {title} {Multichannel spectral mode of the aloha up-conversion interferometer},\ }\href {https://doi.org/10.1093/mnras/sty648} {\bibfield  {journal} {\bibinfo  {journal} {Monthly Notices of the Royal Astronomical Society}\ }\textbf {\bibinfo {volume} {477}},\ \bibinfo {pages} {190} (\bibinfo {year} {2018})}\BibitemShut {NoStop}%
\bibitem [{\citenamefont {Szemendera}\ \emph {et~al.}(2016)\citenamefont {Szemendera}, \citenamefont {Darré}, \citenamefont {Baudoin}, \citenamefont {Grossard}, \citenamefont {Delage}, \citenamefont {Herrmann}, \citenamefont {Silberhorn},\ and\ \citenamefont {Reynaud}}]{szemendera2016mnras}%
  \BibitemOpen
  \bibfield  {author} {\bibinfo {author} {\bibfnamefont {L.}~\bibnamefont {Szemendera}}, \bibinfo {author} {\bibfnamefont {P.}~\bibnamefont {Darré}}, \bibinfo {author} {\bibfnamefont {R.}~\bibnamefont {Baudoin}}, \bibinfo {author} {\bibfnamefont {L.}~\bibnamefont {Grossard}}, \bibinfo {author} {\bibfnamefont {L.}~\bibnamefont {Delage}}, \bibinfo {author} {\bibfnamefont {H.}~\bibnamefont {Herrmann}}, \bibinfo {author} {\bibfnamefont {C.}~\bibnamefont {Silberhorn}},\ and\ \bibinfo {author} {\bibfnamefont {F.}~\bibnamefont {Reynaud}},\ }\bibfield  {title} {\bibinfo {title} {In-lab aloha mid-infrared up-conversion interferometer with high fringe contrast @ $\lambda = 3.39\ \mu m$},\ }\href {https://doi.org/10.1093/mnras/stw196} {\bibfield  {journal} {\bibinfo  {journal} {Monthly Notices of the Royal Astronomical Society}\ }\textbf {\bibinfo {volume} {457}},\ \bibinfo {pages} {3115} (\bibinfo {year} {2016})}\BibitemShut {NoStop}%
\bibitem [{\citenamefont {Lehmann}\ \emph {et~al.}(2019)\citenamefont {Lehmann}, \citenamefont {Grossard}, \citenamefont {Delage}, \citenamefont {Reynaud}, \citenamefont {Chauvet}, \citenamefont {Bassignot}, \citenamefont {Martinache}, \citenamefont {Morand}, \citenamefont {Rivet}, \citenamefont {Schmider},\ and\ \citenamefont {Vernet}}]{lehmann2019mnras}%
  \BibitemOpen
  \bibfield  {author} {\bibinfo {author} {\bibfnamefont {L.}~\bibnamefont {Lehmann}}, \bibinfo {author} {\bibfnamefont {L.}~\bibnamefont {Grossard}}, \bibinfo {author} {\bibfnamefont {L.}~\bibnamefont {Delage}}, \bibinfo {author} {\bibfnamefont {F.}~\bibnamefont {Reynaud}}, \bibinfo {author} {\bibfnamefont {M.}~\bibnamefont {Chauvet}}, \bibinfo {author} {\bibfnamefont {F.}~\bibnamefont {Bassignot}}, \bibinfo {author} {\bibfnamefont {F.}~\bibnamefont {Martinache}}, \bibinfo {author} {\bibfnamefont {F.}~\bibnamefont {Morand}}, \bibinfo {author} {\bibfnamefont {J.-P.}\ \bibnamefont {Rivet}}, \bibinfo {author} {\bibfnamefont {F.-X.}\ \bibnamefont {Schmider}},\ and\ \bibinfo {author} {\bibfnamefont {D.}~\bibnamefont {Vernet}},\ }\bibfield  {title} {\bibinfo {title} {Towards a mid-infrared l band up-conversion interferometer: first on-sky sensitivity test on a single arm},\ }\href {https://doi.org/10.1093/mnras/stz729} {\bibfield  {journal} {\bibinfo  {journal} {Monthly Notices of the Royal Astronomical Society}\
  }\textbf {\bibinfo {volume} {485}},\ \bibinfo {pages} {3595} (\bibinfo {year} {2019})}\BibitemShut {NoStop}%
\bibitem [{\citenamefont {Magri}\ \emph {et~al.}(2020)\citenamefont {Magri}, \citenamefont {Lehmann}, \citenamefont {Grossard}, \citenamefont {Delage}, \citenamefont {Reynaud}, \citenamefont {Chauvet}, \citenamefont {Bassignot}, \citenamefont {Krawczyk},\ and\ \citenamefont {Le Duigou}}]{magri2020mnras}%
  \BibitemOpen
  \bibfield  {author} {\bibinfo {author} {\bibfnamefont {J.}~\bibnamefont {Magri}}, \bibinfo {author} {\bibfnamefont {L.}~\bibnamefont {Lehmann}}, \bibinfo {author} {\bibfnamefont {L.}~\bibnamefont {Grossard}}, \bibinfo {author} {\bibfnamefont {L.}~\bibnamefont {Delage}}, \bibinfo {author} {\bibfnamefont {F.}~\bibnamefont {Reynaud}}, \bibinfo {author} {\bibfnamefont {M.}~\bibnamefont {Chauvet}}, \bibinfo {author} {\bibfnamefont {F.}~\bibnamefont {Bassignot}}, \bibinfo {author} {\bibfnamefont {R.}~\bibnamefont {Krawczyk}},\ and\ \bibinfo {author} {\bibfnamefont {J.-M.}\ \bibnamefont {Le Duigou}},\ }\bibfield  {title} {\bibinfo {title} {Influence of the input-stage architecture on the in-laboratory test of a mid-infrared interferometer: application to the aloha up-conversion interferometer in the l band},\ }\href {https://doi.org/10.1093/mnras/staa3283} {\bibfield  {journal} {\bibinfo  {journal} {Monthly Notices of the Royal Astronomical Society}\ }\textbf {\bibinfo {volume} {501}},\ \bibinfo {pages} {531}
  (\bibinfo {year} {2020})}\BibitemShut {NoStop}%
\bibitem [{\citenamefont {Scully}\ and\ \citenamefont {Zubairy}(1999)}]{scully1999book}%
  \BibitemOpen
  \bibfield  {author} {\bibinfo {author} {\bibfnamefont {M.~O.}\ \bibnamefont {Scully}}\ and\ \bibinfo {author} {\bibfnamefont {M.~S.}\ \bibnamefont {Zubairy}},\ }\href@noop {} {\emph {\bibinfo {title} {Quantum optics}}}\ (\bibinfo  {publisher} {Cambridge University Press},\ \bibinfo {address} {New York, NY},\ \bibinfo {year} {1999})\BibitemShut {NoStop}%
\bibitem [{\citenamefont {Born}\ and\ \citenamefont {Wolf}(1987)}]{born1987book}%
  \BibitemOpen
  \bibfield  {author} {\bibinfo {author} {\bibfnamefont {M.}~\bibnamefont {Born}}\ and\ \bibinfo {author} {\bibfnamefont {E.}~\bibnamefont {Wolf}},\ }\href@noop {} {\emph {\bibinfo {title} {Principles of optics: electromagnetic theory of propagation, interference and diffraction of light}}}\ (\bibinfo  {publisher} {Pergamon Press},\ \bibinfo {address} {Oxford, UK},\ \bibinfo {year} {1987})\BibitemShut {NoStop}%
\bibitem [{\citenamefont {Kumar}(1990)}]{kumar1990ol}%
  \BibitemOpen
  \bibfield  {author} {\bibinfo {author} {\bibfnamefont {P.}~\bibnamefont {Kumar}},\ }\bibfield  {title} {\bibinfo {title} {Quantum frequency conversion},\ }\href {https://doi.org/10.1364/OL.15.001476} {\bibfield  {journal} {\bibinfo  {journal} {Optics Letters}\ }\textbf {\bibinfo {volume} {15}},\ \bibinfo {pages} {1476} (\bibinfo {year} {1990})}\BibitemShut {NoStop}%
\bibitem [{\citenamefont {Albota}\ and\ \citenamefont {Wong}(2004)}]{albota2004ol}%
  \BibitemOpen
  \bibfield  {author} {\bibinfo {author} {\bibfnamefont {M.~A.}\ \bibnamefont {Albota}}\ and\ \bibinfo {author} {\bibfnamefont {F.~N.~C.}\ \bibnamefont {Wong}},\ }\bibfield  {title} {\bibinfo {title} {Efficient single-photon counting at 1.55 µm by means of frequency upconversion},\ }\href {https://doi.org/10.1364/OL.29.001449} {\bibfield  {journal} {\bibinfo  {journal} {Optics Letters}\ }\textbf {\bibinfo {volume} {29}},\ \bibinfo {pages} {1449} (\bibinfo {year} {2004})}\BibitemShut {NoStop}%
\bibitem [{\citenamefont {Collaboration}\ \emph {et~al.}(2019)\citenamefont {Collaboration}, \citenamefont {Akiyama}, \citenamefont {Alberdi}, \citenamefont {Alef}, \citenamefont {Asada}, \citenamefont {Azulay}, \citenamefont {Baczko}, \citenamefont {Ball}, \citenamefont {Baloković}, \citenamefont {Barrett}, \citenamefont {Bintley}, \citenamefont {Blackburn}, \citenamefont {Boland}, \citenamefont {Bouman}, \citenamefont {Bower}, \citenamefont {Bremer}, \citenamefont {Brinkerink}, \citenamefont {Brissenden}, \citenamefont {Britzen}, \citenamefont {Broderick}, \citenamefont {Broguiere}, \citenamefont {Bronzwaer}, \citenamefont {Byun}, \citenamefont {Carlstrom}, \citenamefont {Chael}, \citenamefont {kwan Chan}, \citenamefont {Chatterjee}, \citenamefont {Chatterjee}, \citenamefont {Chen}, \citenamefont {Chen}, \citenamefont {Cho}, \citenamefont {Christian}, \citenamefont {Conway}, \citenamefont {Cordes}, \citenamefont {Crew}, \citenamefont {Cui}, \citenamefont {Davelaar}, \citenamefont {Laurentis}, \citenamefont
  {Deane}, \citenamefont {Dempsey}, \citenamefont {Desvignes}, \citenamefont {Dexter}, \citenamefont {Doeleman}, \citenamefont {Eatough}, \citenamefont {Falcke}, \citenamefont {Fish}, \citenamefont {Fomalont}, \citenamefont {Fraga-Encinas}, \citenamefont {Freeman}, \citenamefont {Friberg}, \citenamefont {Fromm}, \citenamefont {Gómez}, \citenamefont {Galison}, \citenamefont {Gammie}, \citenamefont {García}, \citenamefont {Gentaz}, \citenamefont {Georgiev}, \citenamefont {Goddi}, \citenamefont {Gold}, \citenamefont {Gu}, \citenamefont {Gurwell}, \citenamefont {Hada}, \citenamefont {Hecht}, \citenamefont {Hesper}, \citenamefont {Ho}, \citenamefont {Ho}, \citenamefont {Honma}, \citenamefont {Huang}, \citenamefont {Huang}, \citenamefont {Hughes}, \citenamefont {Ikeda}, \citenamefont {Inoue}, \citenamefont {Issaoun}, \citenamefont {James}, \citenamefont {Jannuzi}, \citenamefont {Janssen}, \citenamefont {Jeter}, \citenamefont {Jiang}, \citenamefont {Johnson}, \citenamefont {Jorstad}, \citenamefont {Jung},
  \citenamefont {Karami}, \citenamefont {Karuppusamy}, \citenamefont {Kawashima}, \citenamefont {Keating}, \citenamefont {Kettenis}, \citenamefont {Kim}, \citenamefont {Kim}, \citenamefont {Kim}, \citenamefont {Kino}, \citenamefont {Koay}, \citenamefont {Koch}, \citenamefont {Koyama}, \citenamefont {Kramer}, \citenamefont {Kramer}, \citenamefont {Krichbaum}, \citenamefont {Kuo}, \citenamefont {Lauer}, \citenamefont {Lee}, \citenamefont {Li}, \citenamefont {Li}, \citenamefont {Lindqvist}, \citenamefont {Liu}, \citenamefont {Liuzzo}, \citenamefont {Lo}, \citenamefont {Lobanov}, \citenamefont {Loinard}, \citenamefont {Lonsdale}, \citenamefont {Lu}, \citenamefont {MacDonald}, \citenamefont {Mao}, \citenamefont {Markoff}, \citenamefont {Marrone}, \citenamefont {Marscher}, \citenamefont {Martí-Vidal}, \citenamefont {Matsushita}, \citenamefont {Matthews}, \citenamefont {Medeiros}, \citenamefont {Menten}, \citenamefont {Mizuno}, \citenamefont {Mizuno}, \citenamefont {Moran}, \citenamefont {Moriyama}, \citenamefont
  {Moscibrodzka}, \citenamefont {Müller}, \citenamefont {Nagai}, \citenamefont {Nagar}, \citenamefont {Nakamura}, \citenamefont {Narayan}, \citenamefont {Narayanan}, \citenamefont {Natarajan}, \citenamefont {Neri}, \citenamefont {Ni}, \citenamefont {Noutsos}, \citenamefont {Okino}, \citenamefont {Olivares}, \citenamefont {Ortiz-León}, \citenamefont {Oyama}, \citenamefont {Özel}, \citenamefont {Palumbo}, \citenamefont {Patel}, \citenamefont {Pen}, \citenamefont {Pesce}, \citenamefont {Piétu}, \citenamefont {Plambeck}, \citenamefont {PopStefanija}, \citenamefont {Porth}, \citenamefont {Prather}, \citenamefont {Preciado-López}, \citenamefont {Psaltis}, \citenamefont {Pu}, \citenamefont {Ramakrishnan}, \citenamefont {Rao}, \citenamefont {Rawlings}, \citenamefont {Raymond}, \citenamefont {Rezzolla}, \citenamefont {Ripperda}, \citenamefont {Roelofs}, \citenamefont {Rogers}, \citenamefont {Ros}, \citenamefont {Rose}, \citenamefont {Roshanineshat}, \citenamefont {Rottmann}, \citenamefont {Roy}, \citenamefont
  {Ruszczyk}, \citenamefont {Ryan}, \citenamefont {Rygl}, \citenamefont {Sánchez}, \citenamefont {Sánchez-Arguelles}, \citenamefont {Sasada}, \citenamefont {Savolainen}, \citenamefont {Schloerb}, \citenamefont {Schuster}, \citenamefont {Shao}, \citenamefont {Shen}, \citenamefont {Small}, \citenamefont {Sohn}, \citenamefont {SooHoo}, \citenamefont {Tazaki}, \citenamefont {Tiede}, \citenamefont {Tilanus}, \citenamefont {Titus}, \citenamefont {Toma}, \citenamefont {Torne}, \citenamefont {Trent}, \citenamefont {Trippe}, \citenamefont {Tsuda}, \citenamefont {van Bemmel}, \citenamefont {van Langevelde}, \citenamefont {van Rossum}, \citenamefont {Wagner}, \citenamefont {Wardle}, \citenamefont {Weintroub}, \citenamefont {Wex}, \citenamefont {Wharton}, \citenamefont {Wielgus}, \citenamefont {Wong}, \citenamefont {Wu}, \citenamefont {Young}, \citenamefont {Young}, \citenamefont {Younsi}, \citenamefont {Yuan}, \citenamefont {Yuan}, \citenamefont {Zensus}, \citenamefont {Zhao}, \citenamefont {Zhao}, \citenamefont
  {Zhu}, \citenamefont {Algaba}, \citenamefont {Allardi}, \citenamefont {Amestica}, \citenamefont {Anczarski}, \citenamefont {Bach}, \citenamefont {Baganoff}, \citenamefont {Beaudoin}, \citenamefont {Benson}, \citenamefont {Berthold}, \citenamefont {Blanchard}, \citenamefont {Blundell}, \citenamefont {Bustamente}, \citenamefont {Cappallo}, \citenamefont {Castillo-Domínguez}, \citenamefont {Chang}, \citenamefont {Chang}, \citenamefont {Chang}, \citenamefont {Chen}, \citenamefont {Chilson}, \citenamefont {Chuter}, \citenamefont {Rosado}, \citenamefont {Coulson}, \citenamefont {Crawford}, \citenamefont {Crowley}, \citenamefont {David}, \citenamefont {Derome}, \citenamefont {Dexter}, \citenamefont {Dornbusch}, \citenamefont {Dudevoir}, \citenamefont {Dzib}, \citenamefont {Eckart}, \citenamefont {Eckert}, \citenamefont {Erickson}, \citenamefont {Everett}, \citenamefont {Faber}, \citenamefont {Farah}, \citenamefont {Fath}, \citenamefont {Folkers}, \citenamefont {Forbes}, \citenamefont {Freund}, \citenamefont
  {Gómez-Ruiz}, \citenamefont {Gale}, \citenamefont {Gao}, \citenamefont {Geertsema}, \citenamefont {Graham}, \citenamefont {Greer}, \citenamefont {Grosslein}, \citenamefont {Gueth}, \citenamefont {Haggard}, \citenamefont {Halverson}, \citenamefont {Han}, \citenamefont {Han}, \citenamefont {Hao}, \citenamefont {Hasegawa}, \citenamefont {Henning}, \citenamefont {Hernández-Gómez}, \citenamefont {Herrero-Illana}, \citenamefont {Heyminck}, \citenamefont {Hirota}, \citenamefont {Hoge}, \citenamefont {Huang}, \citenamefont {Impellizzeri}, \citenamefont {Jiang}, \citenamefont {Kamble}, \citenamefont {Keisler}, \citenamefont {Kimura}, \citenamefont {Kono}, \citenamefont {Kubo}, \citenamefont {Kuroda}, \citenamefont {Lacasse}, \citenamefont {Laing}, \citenamefont {Leitch}, \citenamefont {Li}, \citenamefont {Lin}, \citenamefont {Liu}, \citenamefont {Liu}, \citenamefont {Lu}, \citenamefont {Marson}, \citenamefont {Martin-Cocher}, \citenamefont {Massingill}, \citenamefont {Matulonis}, \citenamefont {McColl},
  \citenamefont {McWhirter}, \citenamefont {Messias}, \citenamefont {Meyer-Zhao}, \citenamefont {Michalik}, \citenamefont {Montaña}, \citenamefont {Montgomerie}, \citenamefont {Mora-Klein}, \citenamefont {Muders}, \citenamefont {Nadolski}, \citenamefont {Navarro}, \citenamefont {Neilsen}, \citenamefont {Nguyen}, \citenamefont {Nishioka}, \citenamefont {Norton}, \citenamefont {Nowak}, \citenamefont {Nystrom}, \citenamefont {Ogawa}, \citenamefont {Oshiro}, \citenamefont {Oyama}, \citenamefont {Parsons}, \citenamefont {Paine}, \citenamefont {Peñalver}, \citenamefont {Phillips}, \citenamefont {Poirier}, \citenamefont {Pradel}, \citenamefont {Primiani}, \citenamefont {Raffin}, \citenamefont {Rahlin}, \citenamefont {Reiland}, \citenamefont {Risacher}, \citenamefont {Ruiz}, \citenamefont {Sáez-Madaín}, \citenamefont {Sassella}, \citenamefont {Schellart}, \citenamefont {Shaw}, \citenamefont {Silva}, \citenamefont {Shiokawa}, \citenamefont {Smith}, \citenamefont {Snow}, \citenamefont {Souccar}, \citenamefont
  {Sousa}, \citenamefont {Sridharan}, \citenamefont {Srinivasan}, \citenamefont {Stahm}, \citenamefont {Stark}, \citenamefont {Story}, \citenamefont {Timmer}, \citenamefont {Vertatschitsch}, \citenamefont {Walther}, \citenamefont {Wei}, \citenamefont {Whitehorn}, \citenamefont {Whitney}, \citenamefont {Woody}, \citenamefont {Wouterloot}, \citenamefont {Wright}, \citenamefont {Yamaguchi}, \citenamefont {Yu}, \citenamefont {Zeballos}, \citenamefont {Zhang},\ and\ \citenamefont {Ziurys}}]{EHT2019ApJ}%
  \BibitemOpen
  \bibfield  {author} {\bibinfo {author} {\bibfnamefont {T.~E. H.~T.}\ \bibnamefont {Collaboration}}, \bibinfo {author} {\bibfnamefont {K.}~\bibnamefont {Akiyama}}, \bibinfo {author} {\bibfnamefont {A.}~\bibnamefont {Alberdi}}, \bibinfo {author} {\bibfnamefont {W.}~\bibnamefont {Alef}}, \bibinfo {author} {\bibfnamefont {K.}~\bibnamefont {Asada}}, \bibinfo {author} {\bibfnamefont {R.}~\bibnamefont {Azulay}}, \bibinfo {author} {\bibfnamefont {A.-K.}\ \bibnamefont {Baczko}}, \bibinfo {author} {\bibfnamefont {D.}~\bibnamefont {Ball}}, \bibinfo {author} {\bibfnamefont {M.}~\bibnamefont {Baloković}}, \bibinfo {author} {\bibfnamefont {J.}~\bibnamefont {Barrett}}, \bibinfo {author} {\bibfnamefont {D.}~\bibnamefont {Bintley}}, \bibinfo {author} {\bibfnamefont {L.}~\bibnamefont {Blackburn}}, \bibinfo {author} {\bibfnamefont {W.}~\bibnamefont {Boland}}, \bibinfo {author} {\bibfnamefont {K.~L.}\ \bibnamefont {Bouman}}, \bibinfo {author} {\bibfnamefont {G.~C.}\ \bibnamefont {Bower}}, \bibinfo {author} {\bibfnamefont
  {M.}~\bibnamefont {Bremer}}, \bibinfo {author} {\bibfnamefont {C.~D.}\ \bibnamefont {Brinkerink}}, \bibinfo {author} {\bibfnamefont {R.}~\bibnamefont {Brissenden}}, \bibinfo {author} {\bibfnamefont {S.}~\bibnamefont {Britzen}}, \bibinfo {author} {\bibfnamefont {A.~E.}\ \bibnamefont {Broderick}}, \bibinfo {author} {\bibfnamefont {D.}~\bibnamefont {Broguiere}}, \bibinfo {author} {\bibfnamefont {T.}~\bibnamefont {Bronzwaer}}, \bibinfo {author} {\bibfnamefont {D.-Y.}\ \bibnamefont {Byun}}, \bibinfo {author} {\bibfnamefont {J.~E.}\ \bibnamefont {Carlstrom}}, \bibinfo {author} {\bibfnamefont {A.}~\bibnamefont {Chael}}, \bibinfo {author} {\bibfnamefont {C.}~\bibnamefont {kwan Chan}}, \bibinfo {author} {\bibfnamefont {S.}~\bibnamefont {Chatterjee}}, \bibinfo {author} {\bibfnamefont {K.}~\bibnamefont {Chatterjee}}, \bibinfo {author} {\bibfnamefont {M.-T.}\ \bibnamefont {Chen}}, \bibinfo {author} {\bibfnamefont {Y.}~\bibnamefont {Chen}}, \bibinfo {author} {\bibfnamefont {I.}~\bibnamefont {Cho}}, \bibinfo {author}
  {\bibfnamefont {P.}~\bibnamefont {Christian}}, \bibinfo {author} {\bibfnamefont {J.~E.}\ \bibnamefont {Conway}}, \bibinfo {author} {\bibfnamefont {J.~M.}\ \bibnamefont {Cordes}}, \bibinfo {author} {\bibfnamefont {G.~B.}\ \bibnamefont {Crew}}, \bibinfo {author} {\bibfnamefont {Y.}~\bibnamefont {Cui}}, \bibinfo {author} {\bibfnamefont {J.}~\bibnamefont {Davelaar}}, \bibinfo {author} {\bibfnamefont {M.~D.}\ \bibnamefont {Laurentis}}, \bibinfo {author} {\bibfnamefont {R.}~\bibnamefont {Deane}}, \bibinfo {author} {\bibfnamefont {J.}~\bibnamefont {Dempsey}}, \bibinfo {author} {\bibfnamefont {G.}~\bibnamefont {Desvignes}}, \bibinfo {author} {\bibfnamefont {J.}~\bibnamefont {Dexter}}, \bibinfo {author} {\bibfnamefont {S.~S.}\ \bibnamefont {Doeleman}}, \bibinfo {author} {\bibfnamefont {R.~P.}\ \bibnamefont {Eatough}}, \bibinfo {author} {\bibfnamefont {H.}~\bibnamefont {Falcke}}, \bibinfo {author} {\bibfnamefont {V.~L.}\ \bibnamefont {Fish}}, \bibinfo {author} {\bibfnamefont {E.}~\bibnamefont {Fomalont}}, \bibinfo
  {author} {\bibfnamefont {R.}~\bibnamefont {Fraga-Encinas}}, \bibinfo {author} {\bibfnamefont {W.~T.}\ \bibnamefont {Freeman}}, \bibinfo {author} {\bibfnamefont {P.}~\bibnamefont {Friberg}}, \bibinfo {author} {\bibfnamefont {C.~M.}\ \bibnamefont {Fromm}}, \bibinfo {author} {\bibfnamefont {J.~L.}\ \bibnamefont {Gómez}}, \bibinfo {author} {\bibfnamefont {P.}~\bibnamefont {Galison}}, \bibinfo {author} {\bibfnamefont {C.~F.}\ \bibnamefont {Gammie}}, \bibinfo {author} {\bibfnamefont {R.}~\bibnamefont {García}}, \bibinfo {author} {\bibfnamefont {O.}~\bibnamefont {Gentaz}}, \bibinfo {author} {\bibfnamefont {B.}~\bibnamefont {Georgiev}}, \bibinfo {author} {\bibfnamefont {C.}~\bibnamefont {Goddi}}, \bibinfo {author} {\bibfnamefont {R.}~\bibnamefont {Gold}}, \bibinfo {author} {\bibfnamefont {M.}~\bibnamefont {Gu}}, \bibinfo {author} {\bibfnamefont {M.}~\bibnamefont {Gurwell}}, \bibinfo {author} {\bibfnamefont {K.}~\bibnamefont {Hada}}, \bibinfo {author} {\bibfnamefont {M.~H.}\ \bibnamefont {Hecht}}, \bibinfo
  {author} {\bibfnamefont {R.}~\bibnamefont {Hesper}}, \bibinfo {author} {\bibfnamefont {L.~C.}\ \bibnamefont {Ho}}, \bibinfo {author} {\bibfnamefont {P.}~\bibnamefont {Ho}}, \bibinfo {author} {\bibfnamefont {M.}~\bibnamefont {Honma}}, \bibinfo {author} {\bibfnamefont {C.-W.~L.}\ \bibnamefont {Huang}}, \bibinfo {author} {\bibfnamefont {L.}~\bibnamefont {Huang}}, \bibinfo {author} {\bibfnamefont {D.~H.}\ \bibnamefont {Hughes}}, \bibinfo {author} {\bibfnamefont {S.}~\bibnamefont {Ikeda}}, \bibinfo {author} {\bibfnamefont {M.}~\bibnamefont {Inoue}}, \bibinfo {author} {\bibfnamefont {S.}~\bibnamefont {Issaoun}}, \bibinfo {author} {\bibfnamefont {D.~J.}\ \bibnamefont {James}}, \bibinfo {author} {\bibfnamefont {B.~T.}\ \bibnamefont {Jannuzi}}, \bibinfo {author} {\bibfnamefont {M.}~\bibnamefont {Janssen}}, \bibinfo {author} {\bibfnamefont {B.}~\bibnamefont {Jeter}}, \bibinfo {author} {\bibfnamefont {W.}~\bibnamefont {Jiang}}, \bibinfo {author} {\bibfnamefont {M.~D.}\ \bibnamefont {Johnson}}, \bibinfo {author}
  {\bibfnamefont {S.}~\bibnamefont {Jorstad}}, \bibinfo {author} {\bibfnamefont {T.}~\bibnamefont {Jung}}, \bibinfo {author} {\bibfnamefont {M.}~\bibnamefont {Karami}}, \bibinfo {author} {\bibfnamefont {R.}~\bibnamefont {Karuppusamy}}, \bibinfo {author} {\bibfnamefont {T.}~\bibnamefont {Kawashima}}, \bibinfo {author} {\bibfnamefont {G.~K.}\ \bibnamefont {Keating}}, \bibinfo {author} {\bibfnamefont {M.}~\bibnamefont {Kettenis}}, \bibinfo {author} {\bibfnamefont {J.-Y.}\ \bibnamefont {Kim}}, \bibinfo {author} {\bibfnamefont {J.}~\bibnamefont {Kim}}, \bibinfo {author} {\bibfnamefont {J.}~\bibnamefont {Kim}}, \bibinfo {author} {\bibfnamefont {M.}~\bibnamefont {Kino}}, \bibinfo {author} {\bibfnamefont {J.~Y.}\ \bibnamefont {Koay}}, \bibinfo {author} {\bibfnamefont {P.~M.}\ \bibnamefont {Koch}}, \bibinfo {author} {\bibfnamefont {S.}~\bibnamefont {Koyama}}, \bibinfo {author} {\bibfnamefont {M.}~\bibnamefont {Kramer}}, \bibinfo {author} {\bibfnamefont {C.}~\bibnamefont {Kramer}}, \bibinfo {author} {\bibfnamefont
  {T.~P.}\ \bibnamefont {Krichbaum}}, \bibinfo {author} {\bibfnamefont {C.-Y.}\ \bibnamefont {Kuo}}, \bibinfo {author} {\bibfnamefont {T.~R.}\ \bibnamefont {Lauer}}, \bibinfo {author} {\bibfnamefont {S.-S.}\ \bibnamefont {Lee}}, \bibinfo {author} {\bibfnamefont {Y.-R.}\ \bibnamefont {Li}}, \bibinfo {author} {\bibfnamefont {Z.}~\bibnamefont {Li}}, \bibinfo {author} {\bibfnamefont {M.}~\bibnamefont {Lindqvist}}, \bibinfo {author} {\bibfnamefont {K.}~\bibnamefont {Liu}}, \bibinfo {author} {\bibfnamefont {E.}~\bibnamefont {Liuzzo}}, \bibinfo {author} {\bibfnamefont {W.-P.}\ \bibnamefont {Lo}}, \bibinfo {author} {\bibfnamefont {A.~P.}\ \bibnamefont {Lobanov}}, \bibinfo {author} {\bibfnamefont {L.}~\bibnamefont {Loinard}}, \bibinfo {author} {\bibfnamefont {C.}~\bibnamefont {Lonsdale}}, \bibinfo {author} {\bibfnamefont {R.-S.}\ \bibnamefont {Lu}}, \bibinfo {author} {\bibfnamefont {N.~R.}\ \bibnamefont {MacDonald}}, \bibinfo {author} {\bibfnamefont {J.}~\bibnamefont {Mao}}, \bibinfo {author} {\bibfnamefont
  {S.}~\bibnamefont {Markoff}}, \bibinfo {author} {\bibfnamefont {D.~P.}\ \bibnamefont {Marrone}}, \bibinfo {author} {\bibfnamefont {A.~P.}\ \bibnamefont {Marscher}}, \bibinfo {author} {\bibfnamefont {I.}~\bibnamefont {Martí-Vidal}}, \bibinfo {author} {\bibfnamefont {S.}~\bibnamefont {Matsushita}}, \bibinfo {author} {\bibfnamefont {L.~D.}\ \bibnamefont {Matthews}}, \bibinfo {author} {\bibfnamefont {L.}~\bibnamefont {Medeiros}}, \bibinfo {author} {\bibfnamefont {K.~M.}\ \bibnamefont {Menten}}, \bibinfo {author} {\bibfnamefont {Y.}~\bibnamefont {Mizuno}}, \bibinfo {author} {\bibfnamefont {I.}~\bibnamefont {Mizuno}}, \bibinfo {author} {\bibfnamefont {J.~M.}\ \bibnamefont {Moran}}, \bibinfo {author} {\bibfnamefont {K.}~\bibnamefont {Moriyama}}, \bibinfo {author} {\bibfnamefont {M.}~\bibnamefont {Moscibrodzka}}, \bibinfo {author} {\bibfnamefont {C.}~\bibnamefont {Müller}}, \bibinfo {author} {\bibfnamefont {H.}~\bibnamefont {Nagai}}, \bibinfo {author} {\bibfnamefont {N.~M.}\ \bibnamefont {Nagar}}, \bibinfo
  {author} {\bibfnamefont {M.}~\bibnamefont {Nakamura}}, \bibinfo {author} {\bibfnamefont {R.}~\bibnamefont {Narayan}}, \bibinfo {author} {\bibfnamefont {G.}~\bibnamefont {Narayanan}}, \bibinfo {author} {\bibfnamefont {I.}~\bibnamefont {Natarajan}}, \bibinfo {author} {\bibfnamefont {R.}~\bibnamefont {Neri}}, \bibinfo {author} {\bibfnamefont {C.}~\bibnamefont {Ni}}, \bibinfo {author} {\bibfnamefont {A.}~\bibnamefont {Noutsos}}, \bibinfo {author} {\bibfnamefont {H.}~\bibnamefont {Okino}}, \bibinfo {author} {\bibfnamefont {H.}~\bibnamefont {Olivares}}, \bibinfo {author} {\bibfnamefont {G.~N.}\ \bibnamefont {Ortiz-León}}, \bibinfo {author} {\bibfnamefont {T.}~\bibnamefont {Oyama}}, \bibinfo {author} {\bibfnamefont {F.}~\bibnamefont {Özel}}, \bibinfo {author} {\bibfnamefont {D.~C.~M.}\ \bibnamefont {Palumbo}}, \bibinfo {author} {\bibfnamefont {N.}~\bibnamefont {Patel}}, \bibinfo {author} {\bibfnamefont {U.-L.}\ \bibnamefont {Pen}}, \bibinfo {author} {\bibfnamefont {D.~W.}\ \bibnamefont {Pesce}}, \bibinfo
  {author} {\bibfnamefont {V.}~\bibnamefont {Piétu}}, \bibinfo {author} {\bibfnamefont {R.}~\bibnamefont {Plambeck}}, \bibinfo {author} {\bibfnamefont {A.}~\bibnamefont {PopStefanija}}, \bibinfo {author} {\bibfnamefont {O.}~\bibnamefont {Porth}}, \bibinfo {author} {\bibfnamefont {B.}~\bibnamefont {Prather}}, \bibinfo {author} {\bibfnamefont {J.~A.}\ \bibnamefont {Preciado-López}}, \bibinfo {author} {\bibfnamefont {D.}~\bibnamefont {Psaltis}}, \bibinfo {author} {\bibfnamefont {H.-Y.}\ \bibnamefont {Pu}}, \bibinfo {author} {\bibfnamefont {V.}~\bibnamefont {Ramakrishnan}}, \bibinfo {author} {\bibfnamefont {R.}~\bibnamefont {Rao}}, \bibinfo {author} {\bibfnamefont {M.~G.}\ \bibnamefont {Rawlings}}, \bibinfo {author} {\bibfnamefont {A.~W.}\ \bibnamefont {Raymond}}, \bibinfo {author} {\bibfnamefont {L.}~\bibnamefont {Rezzolla}}, \bibinfo {author} {\bibfnamefont {B.}~\bibnamefont {Ripperda}}, \bibinfo {author} {\bibfnamefont {F.}~\bibnamefont {Roelofs}}, \bibinfo {author} {\bibfnamefont {A.}~\bibnamefont
  {Rogers}}, \bibinfo {author} {\bibfnamefont {E.}~\bibnamefont {Ros}}, \bibinfo {author} {\bibfnamefont {M.}~\bibnamefont {Rose}}, \bibinfo {author} {\bibfnamefont {A.}~\bibnamefont {Roshanineshat}}, \bibinfo {author} {\bibfnamefont {H.}~\bibnamefont {Rottmann}}, \bibinfo {author} {\bibfnamefont {A.~L.}\ \bibnamefont {Roy}}, \bibinfo {author} {\bibfnamefont {C.}~\bibnamefont {Ruszczyk}}, \bibinfo {author} {\bibfnamefont {B.~R.}\ \bibnamefont {Ryan}}, \bibinfo {author} {\bibfnamefont {K.~L.~J.}\ \bibnamefont {Rygl}}, \bibinfo {author} {\bibfnamefont {S.}~\bibnamefont {Sánchez}}, \bibinfo {author} {\bibfnamefont {D.}~\bibnamefont {Sánchez-Arguelles}}, \bibinfo {author} {\bibfnamefont {M.}~\bibnamefont {Sasada}}, \bibinfo {author} {\bibfnamefont {T.}~\bibnamefont {Savolainen}}, \bibinfo {author} {\bibfnamefont {F.~P.}\ \bibnamefont {Schloerb}}, \bibinfo {author} {\bibfnamefont {K.-F.}\ \bibnamefont {Schuster}}, \bibinfo {author} {\bibfnamefont {L.}~\bibnamefont {Shao}}, \bibinfo {author} {\bibfnamefont
  {Z.}~\bibnamefont {Shen}}, \bibinfo {author} {\bibfnamefont {D.}~\bibnamefont {Small}}, \bibinfo {author} {\bibfnamefont {B.~W.}\ \bibnamefont {Sohn}}, \bibinfo {author} {\bibfnamefont {J.}~\bibnamefont {SooHoo}}, \bibinfo {author} {\bibfnamefont {F.}~\bibnamefont {Tazaki}}, \bibinfo {author} {\bibfnamefont {P.}~\bibnamefont {Tiede}}, \bibinfo {author} {\bibfnamefont {R.~P.~J.}\ \bibnamefont {Tilanus}}, \bibinfo {author} {\bibfnamefont {M.}~\bibnamefont {Titus}}, \bibinfo {author} {\bibfnamefont {K.}~\bibnamefont {Toma}}, \bibinfo {author} {\bibfnamefont {P.}~\bibnamefont {Torne}}, \bibinfo {author} {\bibfnamefont {T.}~\bibnamefont {Trent}}, \bibinfo {author} {\bibfnamefont {S.}~\bibnamefont {Trippe}}, \bibinfo {author} {\bibfnamefont {S.}~\bibnamefont {Tsuda}}, \bibinfo {author} {\bibfnamefont {I.}~\bibnamefont {van Bemmel}}, \bibinfo {author} {\bibfnamefont {H.~J.}\ \bibnamefont {van Langevelde}}, \bibinfo {author} {\bibfnamefont {D.~R.}\ \bibnamefont {van Rossum}}, \bibinfo {author} {\bibfnamefont
  {J.}~\bibnamefont {Wagner}}, \bibinfo {author} {\bibfnamefont {J.}~\bibnamefont {Wardle}}, \bibinfo {author} {\bibfnamefont {J.}~\bibnamefont {Weintroub}}, \bibinfo {author} {\bibfnamefont {N.}~\bibnamefont {Wex}}, \bibinfo {author} {\bibfnamefont {R.}~\bibnamefont {Wharton}}, \bibinfo {author} {\bibfnamefont {M.}~\bibnamefont {Wielgus}}, \bibinfo {author} {\bibfnamefont {G.~N.}\ \bibnamefont {Wong}}, \bibinfo {author} {\bibfnamefont {Q.}~\bibnamefont {Wu}}, \bibinfo {author} {\bibfnamefont {K.}~\bibnamefont {Young}}, \bibinfo {author} {\bibfnamefont {A.}~\bibnamefont {Young}}, \bibinfo {author} {\bibfnamefont {Z.}~\bibnamefont {Younsi}}, \bibinfo {author} {\bibfnamefont {F.}~\bibnamefont {Yuan}}, \bibinfo {author} {\bibfnamefont {Y.-F.}\ \bibnamefont {Yuan}}, \bibinfo {author} {\bibfnamefont {J.~A.}\ \bibnamefont {Zensus}}, \bibinfo {author} {\bibfnamefont {G.}~\bibnamefont {Zhao}}, \bibinfo {author} {\bibfnamefont {S.-S.}\ \bibnamefont {Zhao}}, \bibinfo {author} {\bibfnamefont {Z.}~\bibnamefont {Zhu}},
  \bibinfo {author} {\bibfnamefont {J.-C.}\ \bibnamefont {Algaba}}, \bibinfo {author} {\bibfnamefont {A.}~\bibnamefont {Allardi}}, \bibinfo {author} {\bibfnamefont {R.}~\bibnamefont {Amestica}}, \bibinfo {author} {\bibfnamefont {J.}~\bibnamefont {Anczarski}}, \bibinfo {author} {\bibfnamefont {U.}~\bibnamefont {Bach}}, \bibinfo {author} {\bibfnamefont {F.~K.}\ \bibnamefont {Baganoff}}, \bibinfo {author} {\bibfnamefont {C.}~\bibnamefont {Beaudoin}}, \bibinfo {author} {\bibfnamefont {B.~A.}\ \bibnamefont {Benson}}, \bibinfo {author} {\bibfnamefont {R.}~\bibnamefont {Berthold}}, \bibinfo {author} {\bibfnamefont {J.~M.}\ \bibnamefont {Blanchard}}, \bibinfo {author} {\bibfnamefont {R.}~\bibnamefont {Blundell}}, \bibinfo {author} {\bibfnamefont {S.}~\bibnamefont {Bustamente}}, \bibinfo {author} {\bibfnamefont {R.}~\bibnamefont {Cappallo}}, \bibinfo {author} {\bibfnamefont {E.}~\bibnamefont {Castillo-Domínguez}}, \bibinfo {author} {\bibfnamefont {C.-C.}\ \bibnamefont {Chang}}, \bibinfo {author} {\bibfnamefont
  {S.-H.}\ \bibnamefont {Chang}}, \bibinfo {author} {\bibfnamefont {S.-C.}\ \bibnamefont {Chang}}, \bibinfo {author} {\bibfnamefont {C.-C.}\ \bibnamefont {Chen}}, \bibinfo {author} {\bibfnamefont {R.}~\bibnamefont {Chilson}}, \bibinfo {author} {\bibfnamefont {T.~C.}\ \bibnamefont {Chuter}}, \bibinfo {author} {\bibfnamefont {R.~C.}\ \bibnamefont {Rosado}}, \bibinfo {author} {\bibfnamefont {I.~M.}\ \bibnamefont {Coulson}}, \bibinfo {author} {\bibfnamefont {T.~M.}\ \bibnamefont {Crawford}}, \bibinfo {author} {\bibfnamefont {J.}~\bibnamefont {Crowley}}, \bibinfo {author} {\bibfnamefont {J.}~\bibnamefont {David}}, \bibinfo {author} {\bibfnamefont {M.}~\bibnamefont {Derome}}, \bibinfo {author} {\bibfnamefont {M.}~\bibnamefont {Dexter}}, \bibinfo {author} {\bibfnamefont {S.}~\bibnamefont {Dornbusch}}, \bibinfo {author} {\bibfnamefont {K.~A.}\ \bibnamefont {Dudevoir}}, \bibinfo {author} {\bibfnamefont {S.~A.}\ \bibnamefont {Dzib}}, \bibinfo {author} {\bibfnamefont {A.}~\bibnamefont {Eckart}}, \bibinfo {author}
  {\bibfnamefont {C.}~\bibnamefont {Eckert}}, \bibinfo {author} {\bibfnamefont {N.~R.}\ \bibnamefont {Erickson}}, \bibinfo {author} {\bibfnamefont {W.~B.}\ \bibnamefont {Everett}}, \bibinfo {author} {\bibfnamefont {A.}~\bibnamefont {Faber}}, \bibinfo {author} {\bibfnamefont {J.~R.}\ \bibnamefont {Farah}}, \bibinfo {author} {\bibfnamefont {V.}~\bibnamefont {Fath}}, \bibinfo {author} {\bibfnamefont {T.~W.}\ \bibnamefont {Folkers}}, \bibinfo {author} {\bibfnamefont {D.~C.}\ \bibnamefont {Forbes}}, \bibinfo {author} {\bibfnamefont {R.}~\bibnamefont {Freund}}, \bibinfo {author} {\bibfnamefont {A.~I.}\ \bibnamefont {Gómez-Ruiz}}, \bibinfo {author} {\bibfnamefont {D.~M.}\ \bibnamefont {Gale}}, \bibinfo {author} {\bibfnamefont {F.}~\bibnamefont {Gao}}, \bibinfo {author} {\bibfnamefont {G.}~\bibnamefont {Geertsema}}, \bibinfo {author} {\bibfnamefont {D.~A.}\ \bibnamefont {Graham}}, \bibinfo {author} {\bibfnamefont {C.~H.}\ \bibnamefont {Greer}}, \bibinfo {author} {\bibfnamefont {R.}~\bibnamefont {Grosslein}},
  \bibinfo {author} {\bibfnamefont {F.}~\bibnamefont {Gueth}}, \bibinfo {author} {\bibfnamefont {D.}~\bibnamefont {Haggard}}, \bibinfo {author} {\bibfnamefont {N.~W.}\ \bibnamefont {Halverson}}, \bibinfo {author} {\bibfnamefont {C.-C.}\ \bibnamefont {Han}}, \bibinfo {author} {\bibfnamefont {K.-C.}\ \bibnamefont {Han}}, \bibinfo {author} {\bibfnamefont {J.}~\bibnamefont {Hao}}, \bibinfo {author} {\bibfnamefont {Y.}~\bibnamefont {Hasegawa}}, \bibinfo {author} {\bibfnamefont {J.~W.}\ \bibnamefont {Henning}}, \bibinfo {author} {\bibfnamefont {A.}~\bibnamefont {Hernández-Gómez}}, \bibinfo {author} {\bibfnamefont {R.}~\bibnamefont {Herrero-Illana}}, \bibinfo {author} {\bibfnamefont {S.}~\bibnamefont {Heyminck}}, \bibinfo {author} {\bibfnamefont {A.}~\bibnamefont {Hirota}}, \bibinfo {author} {\bibfnamefont {J.}~\bibnamefont {Hoge}}, \bibinfo {author} {\bibfnamefont {Y.-D.}\ \bibnamefont {Huang}}, \bibinfo {author} {\bibfnamefont {C.~M.~V.}\ \bibnamefont {Impellizzeri}}, \bibinfo {author} {\bibfnamefont
  {H.}~\bibnamefont {Jiang}}, \bibinfo {author} {\bibfnamefont {A.}~\bibnamefont {Kamble}}, \bibinfo {author} {\bibfnamefont {R.}~\bibnamefont {Keisler}}, \bibinfo {author} {\bibfnamefont {K.}~\bibnamefont {Kimura}}, \bibinfo {author} {\bibfnamefont {Y.}~\bibnamefont {Kono}}, \bibinfo {author} {\bibfnamefont {D.}~\bibnamefont {Kubo}}, \bibinfo {author} {\bibfnamefont {J.}~\bibnamefont {Kuroda}}, \bibinfo {author} {\bibfnamefont {R.}~\bibnamefont {Lacasse}}, \bibinfo {author} {\bibfnamefont {R.~A.}\ \bibnamefont {Laing}}, \bibinfo {author} {\bibfnamefont {E.~M.}\ \bibnamefont {Leitch}}, \bibinfo {author} {\bibfnamefont {C.-T.}\ \bibnamefont {Li}}, \bibinfo {author} {\bibfnamefont {L.~C.-C.}\ \bibnamefont {Lin}}, \bibinfo {author} {\bibfnamefont {C.-T.}\ \bibnamefont {Liu}}, \bibinfo {author} {\bibfnamefont {K.-Y.}\ \bibnamefont {Liu}}, \bibinfo {author} {\bibfnamefont {L.-M.}\ \bibnamefont {Lu}}, \bibinfo {author} {\bibfnamefont {R.~G.}\ \bibnamefont {Marson}}, \bibinfo {author} {\bibfnamefont {P.~L.}\
  \bibnamefont {Martin-Cocher}}, \bibinfo {author} {\bibfnamefont {K.~D.}\ \bibnamefont {Massingill}}, \bibinfo {author} {\bibfnamefont {C.}~\bibnamefont {Matulonis}}, \bibinfo {author} {\bibfnamefont {M.~P.}\ \bibnamefont {McColl}}, \bibinfo {author} {\bibfnamefont {S.~R.}\ \bibnamefont {McWhirter}}, \bibinfo {author} {\bibfnamefont {H.}~\bibnamefont {Messias}}, \bibinfo {author} {\bibfnamefont {Z.}~\bibnamefont {Meyer-Zhao}}, \bibinfo {author} {\bibfnamefont {D.}~\bibnamefont {Michalik}}, \bibinfo {author} {\bibfnamefont {A.}~\bibnamefont {Montaña}}, \bibinfo {author} {\bibfnamefont {W.}~\bibnamefont {Montgomerie}}, \bibinfo {author} {\bibfnamefont {M.}~\bibnamefont {Mora-Klein}}, \bibinfo {author} {\bibfnamefont {D.}~\bibnamefont {Muders}}, \bibinfo {author} {\bibfnamefont {A.}~\bibnamefont {Nadolski}}, \bibinfo {author} {\bibfnamefont {S.}~\bibnamefont {Navarro}}, \bibinfo {author} {\bibfnamefont {J.}~\bibnamefont {Neilsen}}, \bibinfo {author} {\bibfnamefont {C.~H.}\ \bibnamefont {Nguyen}}, \bibinfo
  {author} {\bibfnamefont {H.}~\bibnamefont {Nishioka}}, \bibinfo {author} {\bibfnamefont {T.}~\bibnamefont {Norton}}, \bibinfo {author} {\bibfnamefont {M.~A.}\ \bibnamefont {Nowak}}, \bibinfo {author} {\bibfnamefont {G.}~\bibnamefont {Nystrom}}, \bibinfo {author} {\bibfnamefont {H.}~\bibnamefont {Ogawa}}, \bibinfo {author} {\bibfnamefont {P.}~\bibnamefont {Oshiro}}, \bibinfo {author} {\bibfnamefont {T.}~\bibnamefont {Oyama}}, \bibinfo {author} {\bibfnamefont {H.}~\bibnamefont {Parsons}}, \bibinfo {author} {\bibfnamefont {S.~N.}\ \bibnamefont {Paine}}, \bibinfo {author} {\bibfnamefont {J.}~\bibnamefont {Peñalver}}, \bibinfo {author} {\bibfnamefont {N.~M.}\ \bibnamefont {Phillips}}, \bibinfo {author} {\bibfnamefont {M.}~\bibnamefont {Poirier}}, \bibinfo {author} {\bibfnamefont {N.}~\bibnamefont {Pradel}}, \bibinfo {author} {\bibfnamefont {R.~A.}\ \bibnamefont {Primiani}}, \bibinfo {author} {\bibfnamefont {P.~A.}\ \bibnamefont {Raffin}}, \bibinfo {author} {\bibfnamefont {A.~S.}\ \bibnamefont {Rahlin}},
  \bibinfo {author} {\bibfnamefont {G.}~\bibnamefont {Reiland}}, \bibinfo {author} {\bibfnamefont {C.}~\bibnamefont {Risacher}}, \bibinfo {author} {\bibfnamefont {I.}~\bibnamefont {Ruiz}}, \bibinfo {author} {\bibfnamefont {A.~F.}\ \bibnamefont {Sáez-Madaín}}, \bibinfo {author} {\bibfnamefont {R.}~\bibnamefont {Sassella}}, \bibinfo {author} {\bibfnamefont {P.}~\bibnamefont {Schellart}}, \bibinfo {author} {\bibfnamefont {P.}~\bibnamefont {Shaw}}, \bibinfo {author} {\bibfnamefont {K.~M.}\ \bibnamefont {Silva}}, \bibinfo {author} {\bibfnamefont {H.}~\bibnamefont {Shiokawa}}, \bibinfo {author} {\bibfnamefont {D.~R.}\ \bibnamefont {Smith}}, \bibinfo {author} {\bibfnamefont {W.}~\bibnamefont {Snow}}, \bibinfo {author} {\bibfnamefont {K.}~\bibnamefont {Souccar}}, \bibinfo {author} {\bibfnamefont {D.}~\bibnamefont {Sousa}}, \bibinfo {author} {\bibfnamefont {T.~K.}\ \bibnamefont {Sridharan}}, \bibinfo {author} {\bibfnamefont {R.}~\bibnamefont {Srinivasan}}, \bibinfo {author} {\bibfnamefont {W.}~\bibnamefont {Stahm}},
  \bibinfo {author} {\bibfnamefont {A.~A.}\ \bibnamefont {Stark}}, \bibinfo {author} {\bibfnamefont {K.}~\bibnamefont {Story}}, \bibinfo {author} {\bibfnamefont {S.~T.}\ \bibnamefont {Timmer}}, \bibinfo {author} {\bibfnamefont {L.}~\bibnamefont {Vertatschitsch}}, \bibinfo {author} {\bibfnamefont {C.}~\bibnamefont {Walther}}, \bibinfo {author} {\bibfnamefont {T.-S.}\ \bibnamefont {Wei}}, \bibinfo {author} {\bibfnamefont {N.}~\bibnamefont {Whitehorn}}, \bibinfo {author} {\bibfnamefont {A.~R.}\ \bibnamefont {Whitney}}, \bibinfo {author} {\bibfnamefont {D.~P.}\ \bibnamefont {Woody}}, \bibinfo {author} {\bibfnamefont {J.~G.~A.}\ \bibnamefont {Wouterloot}}, \bibinfo {author} {\bibfnamefont {M.}~\bibnamefont {Wright}}, \bibinfo {author} {\bibfnamefont {P.}~\bibnamefont {Yamaguchi}}, \bibinfo {author} {\bibfnamefont {C.-Y.}\ \bibnamefont {Yu}}, \bibinfo {author} {\bibfnamefont {M.}~\bibnamefont {Zeballos}}, \bibinfo {author} {\bibfnamefont {S.}~\bibnamefont {Zhang}},\ and\ \bibinfo {author} {\bibfnamefont
  {L.}~\bibnamefont {Ziurys}},\ }\bibfield  {title} {\bibinfo {title} {First m87 event horizon telescope results. i. the shadow of the supermassive black hole},\ }\href {https://doi.org/10.3847/2041-8213/ab0ec7} {\bibfield  {journal} {\bibinfo  {journal} {The Astrophysical Journal Letters}\ }\textbf {\bibinfo {volume} {875}},\ \bibinfo {pages} {L1} (\bibinfo {year} {2019})}\BibitemShut {NoStop}%
\bibitem [{\citenamefont {Prieto}\ \emph {et~al.}(2016)\citenamefont {Prieto}, \citenamefont {Fernández-Ontiveros}, \citenamefont {Markoff}, \citenamefont {Espada},\ and\ \citenamefont {González-Martín}}]{Prieto2016MNRAS}%
  \BibitemOpen
  \bibfield  {author} {\bibinfo {author} {\bibfnamefont {M.~A.}\ \bibnamefont {Prieto}}, \bibinfo {author} {\bibfnamefont {J.~A.}\ \bibnamefont {Fernández-Ontiveros}}, \bibinfo {author} {\bibfnamefont {S.}~\bibnamefont {Markoff}}, \bibinfo {author} {\bibfnamefont {D.}~\bibnamefont {Espada}},\ and\ \bibinfo {author} {\bibfnamefont {O.}~\bibnamefont {González-Martín}},\ }\bibfield  {title} {\bibinfo {title} {{The central parsecs of M87: jet emission and an elusive accretion disc}},\ }\href {https://doi.org/10.1093/mnras/stw166} {\bibfield  {journal} {\bibinfo  {journal} {Monthly Notices of the Royal Astronomical Society}\ }\textbf {\bibinfo {volume} {457}},\ \bibinfo {pages} {3801} (\bibinfo {year} {2016})},\ \Eprint {https://arxiv.org/abs/https://academic.oup.com/mnras/article-pdf/457/4/3801/18511244/stw166.pdf} {https://academic.oup.com/mnras/article-pdf/457/4/3801/18511244/stw166.pdf} \BibitemShut {NoStop}%
\bibitem [{\citenamefont {Rieke}\ \emph {et~al.}(2023)\citenamefont {Rieke}, \citenamefont {Kelly}, \citenamefont {Misselt}, \citenamefont {Stansberry}, \citenamefont {Boyer}, \citenamefont {Beatty}, \citenamefont {Egami}, \citenamefont {Florian}, \citenamefont {Greene}, \citenamefont {Hainline}, \citenamefont {Leisenring}, \citenamefont {Roellig}, \citenamefont {Schlawin}, \citenamefont {Sun}, \citenamefont {Tinnin}, \citenamefont {Williams}, \citenamefont {Willmer}, \citenamefont {Wilson}, \citenamefont {Clark}, \citenamefont {Rohrbach}, \citenamefont {Brooks}, \citenamefont {Canipe}, \citenamefont {Correnti}, \citenamefont {DiFelice}, \citenamefont {Gennaro}, \citenamefont {Girard}, \citenamefont {Hartig}, \citenamefont {Hilbert}, \citenamefont {Koekemoer}, \citenamefont {Nikolov}, \citenamefont {Pirzkal}, \citenamefont {Rest}, \citenamefont {Robberto}, \citenamefont {Sunnquist}, \citenamefont {Telfer}, \citenamefont {Wu}, \citenamefont {Ferry}, \citenamefont {Lewis}, \citenamefont {Baum}, \citenamefont
  {Beichman}, \citenamefont {Doyon}, \citenamefont {Dressler}, \citenamefont {Eisenstein}, \citenamefont {Ferrarese}, \citenamefont {Hodapp}, \citenamefont {Horner}, \citenamefont {Jaffe}, \citenamefont {Johnstone}, \citenamefont {Krist}, \citenamefont {Martin}, \citenamefont {McCarthy}, \citenamefont {Meyer}, \citenamefont {Rieke}, \citenamefont {Trauger},\ and\ \citenamefont {Young}}]{Rieke2023PASP}%
  \BibitemOpen
  \bibfield  {author} {\bibinfo {author} {\bibfnamefont {M.~J.}\ \bibnamefont {Rieke}}, \bibinfo {author} {\bibfnamefont {D.~M.}\ \bibnamefont {Kelly}}, \bibinfo {author} {\bibfnamefont {K.}~\bibnamefont {Misselt}}, \bibinfo {author} {\bibfnamefont {J.}~\bibnamefont {Stansberry}}, \bibinfo {author} {\bibfnamefont {M.}~\bibnamefont {Boyer}}, \bibinfo {author} {\bibfnamefont {T.}~\bibnamefont {Beatty}}, \bibinfo {author} {\bibfnamefont {E.}~\bibnamefont {Egami}}, \bibinfo {author} {\bibfnamefont {M.}~\bibnamefont {Florian}}, \bibinfo {author} {\bibfnamefont {T.~P.}\ \bibnamefont {Greene}}, \bibinfo {author} {\bibfnamefont {K.}~\bibnamefont {Hainline}}, \bibinfo {author} {\bibfnamefont {J.}~\bibnamefont {Leisenring}}, \bibinfo {author} {\bibfnamefont {T.}~\bibnamefont {Roellig}}, \bibinfo {author} {\bibfnamefont {E.}~\bibnamefont {Schlawin}}, \bibinfo {author} {\bibfnamefont {F.}~\bibnamefont {Sun}}, \bibinfo {author} {\bibfnamefont {L.}~\bibnamefont {Tinnin}}, \bibinfo {author} {\bibfnamefont {C.~C.}\
  \bibnamefont {Williams}}, \bibinfo {author} {\bibfnamefont {C.~N.~A.}\ \bibnamefont {Willmer}}, \bibinfo {author} {\bibfnamefont {D.}~\bibnamefont {Wilson}}, \bibinfo {author} {\bibfnamefont {C.~R.}\ \bibnamefont {Clark}}, \bibinfo {author} {\bibfnamefont {S.}~\bibnamefont {Rohrbach}}, \bibinfo {author} {\bibfnamefont {B.}~\bibnamefont {Brooks}}, \bibinfo {author} {\bibfnamefont {A.}~\bibnamefont {Canipe}}, \bibinfo {author} {\bibfnamefont {M.}~\bibnamefont {Correnti}}, \bibinfo {author} {\bibfnamefont {A.}~\bibnamefont {DiFelice}}, \bibinfo {author} {\bibfnamefont {M.}~\bibnamefont {Gennaro}}, \bibinfo {author} {\bibfnamefont {J.}~\bibnamefont {Girard}}, \bibinfo {author} {\bibfnamefont {G.}~\bibnamefont {Hartig}}, \bibinfo {author} {\bibfnamefont {B.}~\bibnamefont {Hilbert}}, \bibinfo {author} {\bibfnamefont {A.~M.}\ \bibnamefont {Koekemoer}}, \bibinfo {author} {\bibfnamefont {N.~K.}\ \bibnamefont {Nikolov}}, \bibinfo {author} {\bibfnamefont {N.}~\bibnamefont {Pirzkal}}, \bibinfo {author} {\bibfnamefont
  {A.}~\bibnamefont {Rest}}, \bibinfo {author} {\bibfnamefont {M.}~\bibnamefont {Robberto}}, \bibinfo {author} {\bibfnamefont {B.}~\bibnamefont {Sunnquist}}, \bibinfo {author} {\bibfnamefont {R.}~\bibnamefont {Telfer}}, \bibinfo {author} {\bibfnamefont {C.~R.}\ \bibnamefont {Wu}}, \bibinfo {author} {\bibfnamefont {M.}~\bibnamefont {Ferry}}, \bibinfo {author} {\bibfnamefont {D.}~\bibnamefont {Lewis}}, \bibinfo {author} {\bibfnamefont {S.}~\bibnamefont {Baum}}, \bibinfo {author} {\bibfnamefont {C.}~\bibnamefont {Beichman}}, \bibinfo {author} {\bibfnamefont {R.}~\bibnamefont {Doyon}}, \bibinfo {author} {\bibfnamefont {A.}~\bibnamefont {Dressler}}, \bibinfo {author} {\bibfnamefont {D.~J.}\ \bibnamefont {Eisenstein}}, \bibinfo {author} {\bibfnamefont {L.}~\bibnamefont {Ferrarese}}, \bibinfo {author} {\bibfnamefont {K.}~\bibnamefont {Hodapp}}, \bibinfo {author} {\bibfnamefont {S.}~\bibnamefont {Horner}}, \bibinfo {author} {\bibfnamefont {D.~T.}\ \bibnamefont {Jaffe}}, \bibinfo {author} {\bibfnamefont
  {D.}~\bibnamefont {Johnstone}}, \bibinfo {author} {\bibfnamefont {J.}~\bibnamefont {Krist}}, \bibinfo {author} {\bibfnamefont {P.}~\bibnamefont {Martin}}, \bibinfo {author} {\bibfnamefont {D.~W.}\ \bibnamefont {McCarthy}}, \bibinfo {author} {\bibfnamefont {M.}~\bibnamefont {Meyer}}, \bibinfo {author} {\bibfnamefont {G.~H.}\ \bibnamefont {Rieke}}, \bibinfo {author} {\bibfnamefont {J.}~\bibnamefont {Trauger}},\ and\ \bibinfo {author} {\bibfnamefont {E.~T.}\ \bibnamefont {Young}},\ }\bibfield  {title} {\bibinfo {title} {Performance of nircam on jwst in flight},\ }\href {https://doi.org/10.1088/1538-3873/acac53} {\bibfield  {journal} {\bibinfo  {journal} {Publications of the Astronomical Society of the Pacific}\ }\textbf {\bibinfo {volume} {135}},\ \bibinfo {pages} {028001} (\bibinfo {year} {2023})}\BibitemShut {NoStop}%
\bibitem [{\citenamefont {Wright}\ \emph {et~al.}(2010)\citenamefont {Wright}, \citenamefont {Eisenhardt}, \citenamefont {Mainzer}, \citenamefont {Ressler}, \citenamefont {Cutri}, \citenamefont {Jarrett}, \citenamefont {Kirkpatrick}, \citenamefont {Padgett}, \citenamefont {McMillan}, \citenamefont {Skrutskie}, \citenamefont {Stanford}, \citenamefont {Cohen}, \citenamefont {Walker}, \citenamefont {Mather}, \citenamefont {Leisawitz}, \citenamefont {Gautier}, \citenamefont {McLean}, \citenamefont {Benford}, \citenamefont {Lonsdale}, \citenamefont {Blain}, \citenamefont {Mendez}, \citenamefont {Irace}, \citenamefont {Duval}, \citenamefont {Liu}, \citenamefont {Royer}, \citenamefont {Heinrichsen}, \citenamefont {Howard}, \citenamefont {Shannon}, \citenamefont {Kendall}, \citenamefont {Walsh}, \citenamefont {Larsen}, \citenamefont {Cardon}, \citenamefont {Schick}, \citenamefont {Schwalm}, \citenamefont {Abid}, \citenamefont {Fabinsky}, \citenamefont {Naes},\ and\ \citenamefont {Tsai}}]{Wright2010AJ}%
  \BibitemOpen
  \bibfield  {author} {\bibinfo {author} {\bibfnamefont {E.~L.}\ \bibnamefont {Wright}}, \bibinfo {author} {\bibfnamefont {P.~R.~M.}\ \bibnamefont {Eisenhardt}}, \bibinfo {author} {\bibfnamefont {A.~K.}\ \bibnamefont {Mainzer}}, \bibinfo {author} {\bibfnamefont {M.~E.}\ \bibnamefont {Ressler}}, \bibinfo {author} {\bibfnamefont {R.~M.}\ \bibnamefont {Cutri}}, \bibinfo {author} {\bibfnamefont {T.}~\bibnamefont {Jarrett}}, \bibinfo {author} {\bibfnamefont {J.~D.}\ \bibnamefont {Kirkpatrick}}, \bibinfo {author} {\bibfnamefont {D.}~\bibnamefont {Padgett}}, \bibinfo {author} {\bibfnamefont {R.~S.}\ \bibnamefont {McMillan}}, \bibinfo {author} {\bibfnamefont {M.}~\bibnamefont {Skrutskie}}, \bibinfo {author} {\bibfnamefont {S.~A.}\ \bibnamefont {Stanford}}, \bibinfo {author} {\bibfnamefont {M.}~\bibnamefont {Cohen}}, \bibinfo {author} {\bibfnamefont {R.~G.}\ \bibnamefont {Walker}}, \bibinfo {author} {\bibfnamefont {J.~C.}\ \bibnamefont {Mather}}, \bibinfo {author} {\bibfnamefont {D.}~\bibnamefont {Leisawitz}},
  \bibinfo {author} {\bibfnamefont {T.~N.}\ \bibnamefont {Gautier}}, \bibinfo {author} {\bibfnamefont {I.}~\bibnamefont {McLean}}, \bibinfo {author} {\bibfnamefont {D.}~\bibnamefont {Benford}}, \bibinfo {author} {\bibfnamefont {C.~J.}\ \bibnamefont {Lonsdale}}, \bibinfo {author} {\bibfnamefont {A.}~\bibnamefont {Blain}}, \bibinfo {author} {\bibfnamefont {B.}~\bibnamefont {Mendez}}, \bibinfo {author} {\bibfnamefont {W.~R.}\ \bibnamefont {Irace}}, \bibinfo {author} {\bibfnamefont {V.}~\bibnamefont {Duval}}, \bibinfo {author} {\bibfnamefont {F.}~\bibnamefont {Liu}}, \bibinfo {author} {\bibfnamefont {D.}~\bibnamefont {Royer}}, \bibinfo {author} {\bibfnamefont {I.}~\bibnamefont {Heinrichsen}}, \bibinfo {author} {\bibfnamefont {J.}~\bibnamefont {Howard}}, \bibinfo {author} {\bibfnamefont {M.}~\bibnamefont {Shannon}}, \bibinfo {author} {\bibfnamefont {M.}~\bibnamefont {Kendall}}, \bibinfo {author} {\bibfnamefont {A.~L.}\ \bibnamefont {Walsh}}, \bibinfo {author} {\bibfnamefont {M.}~\bibnamefont {Larsen}}, \bibinfo
  {author} {\bibfnamefont {J.~G.}\ \bibnamefont {Cardon}}, \bibinfo {author} {\bibfnamefont {S.}~\bibnamefont {Schick}}, \bibinfo {author} {\bibfnamefont {M.}~\bibnamefont {Schwalm}}, \bibinfo {author} {\bibfnamefont {M.}~\bibnamefont {Abid}}, \bibinfo {author} {\bibfnamefont {B.}~\bibnamefont {Fabinsky}}, \bibinfo {author} {\bibfnamefont {L.}~\bibnamefont {Naes}},\ and\ \bibinfo {author} {\bibfnamefont {C.-W.}\ \bibnamefont {Tsai}},\ }\bibfield  {title} {\bibinfo {title} {The wide-field infrared survey explorer (wise): Mission description and initial on-orbit performance},\ }\href {https://doi.org/10.1088/0004-6256/140/6/1868} {\bibfield  {journal} {\bibinfo  {journal} {The Astronomical Journal}\ }\textbf {\bibinfo {volume} {140}},\ \bibinfo {pages} {1868} (\bibinfo {year} {2010})}\BibitemShut {NoStop}%
\bibitem [{\citenamefont {Ge}\ \emph {et~al.}(2024)\citenamefont {Ge}, \citenamefont {Han}, \citenamefont {Yang}, \citenamefont {Wang}, \citenamefont {Li}, \citenamefont {Li}, \citenamefont {Gao}, \citenamefont {Chen}, \citenamefont {Niu}, \citenamefont {Xie}, \citenamefont {Zhou},\ and\ \citenamefont {Shi}}]{Ge2024SciAdv}%
  \BibitemOpen
  \bibfield  {author} {\bibinfo {author} {\bibfnamefont {Z.}~\bibnamefont {Ge}}, \bibinfo {author} {\bibfnamefont {Z.-Q.-Z.}\ \bibnamefont {Han}}, \bibinfo {author} {\bibfnamefont {F.}~\bibnamefont {Yang}}, \bibinfo {author} {\bibfnamefont {X.-H.}\ \bibnamefont {Wang}}, \bibinfo {author} {\bibfnamefont {Y.-H.}\ \bibnamefont {Li}}, \bibinfo {author} {\bibfnamefont {Y.}~\bibnamefont {Li}}, \bibinfo {author} {\bibfnamefont {M.-Y.}\ \bibnamefont {Gao}}, \bibinfo {author} {\bibfnamefont {R.-H.}\ \bibnamefont {Chen}}, \bibinfo {author} {\bibfnamefont {S.-J.}\ \bibnamefont {Niu}}, \bibinfo {author} {\bibfnamefont {M.-Y.}\ \bibnamefont {Xie}}, \bibinfo {author} {\bibfnamefont {Z.-Y.}\ \bibnamefont {Zhou}},\ and\ \bibinfo {author} {\bibfnamefont {B.-S.}\ \bibnamefont {Shi}},\ }\bibfield  {title} {\bibinfo {title} {Quantum entanglement and interference at 3 $\mu m$},\ }\href {https://doi.org/10.1126/sciadv.adm7565} {\bibfield  {journal} {\bibinfo  {journal} {Science Advances}\ }\textbf {\bibinfo {volume} {10}},\
  \bibinfo {pages} {eadm7565} (\bibinfo {year} {2024})}\BibitemShut {NoStop}%
\end{thebibliography}%

\end{document}